\documentclass[a4paper,prd,aps,floats,preprintnumbers,showpacs,twocolumn,amsmath,amssymb]{revtex4}
\usepackage{graphicx}
\usepackage{placeins}

\begin{document}

\preprint{
\vbox{
\hbox{MUP-06-??/T???}
\hbox{ADP-06-04/T635}
}}

\title{Gluon flux-tube distribution and linear confinement in baryons} 

\author{F. Bissey}
\author{F-G. Cao}
\author{A. R. Kitson}
\author{A. I. Signal}
\affiliation{Institute of Fundamental Sciences, Massey University,\\
        Private Bag 11 222, Palmerston North, New Zealand}
\author{D. B. Leinweber}
\author{B. G. Lasscock}
\author{A. G. Williams}
\affiliation{Centre for the Subatomic Structure of Matter and\\
        Department of Physics, University of Adelaide, SA 5005, Australia}

%\date{May 4, 2006}

\begin{abstract}
The distribution of gluon fields in a baryon is of fundamental
interest in QCD.  We have observed the formation of gluon flux-tubes
within baryons using lattice QCD techniques.  In particular we use a
high-statistics approach, based on the translational and rotational
symmetries of the four-dimensional lattice, which enables us to
observe correlations between the vacuum action density and the quark
positions in a completely gauge independent manner.  This contrasts
with earlier studies which needed to use gauge-dependent smoothing
techniques.  We use 200 ${\cal O}(a^2)$-improved quenched QCD
gauge-field configurations on a $16^3 \times 32$ lattice with a
lattice spacing of 0.123 fm.  Vacuum field fluctuations are observed
to be suppressed in the presence of static quarks such that flux tubes
represent the suppression of gluon-field fluctuations.  We considered
numerous different link paths in the creation of the static quark
sources in order to investigate the dependence of the flux tubes on
the source shape.  We have analyzed 11 {\em L} shapes and 8 {\em T}
and {\em Y} shapes of varying sizes in order to explore a variety of
flux-tube topologies, including the ground state.  At large
separations, Y-shape flux-tube formation is observed.  T-shaped paths
are observed to relax towards a Y-shaped topology, whereas L-shaped
paths give rise to a large potential energy.  We do not find any
evidence for the formation of a $\Delta$-shaped flux-tube (empty
triangle) distribution.  However, at small quark separations, we do
observe an expulsion of gluon-field fluctuations in the shape of a
filled triangle with maximal expulsion at the centre of the
triangle. Having identified the precise geometry of the flux
distribution, we are able to perform a quantitative comparison between
the length of the flux-tube and the associated static quark potential.
For every source configuration considered we find a universal string
tension, and conclude that, for large quark separations, the ground
state potential is that which minimizes the length of the flux-tube.
The characteristic flux tube radius of the baryonic ground state
potential is found to be $0.38 \pm 0.03$ fm, with vacuum fluctuations
suppressed by $7.2 \pm 0.6 \%$.  The node connecting the flux tubes is
25\% larger at $0.47 \pm 0.02$ fm with a larger suppression of the
vacuum by $8.1 \pm 0.7 \%$
\end{abstract}


\maketitle

\section{INTRODUCTION}

Recently there has been renewed interest in studying the distribution
of quark and gluon fields in the three-quark static-baryon system.
While the earliest studies were inconclusive \cite{Flower:1986ru}, improved
computing resources and analysis techniques now make it possible to
study this system in a quantitative manner
\cite{Takahashi:2002bw,Alexandrou:2001ip,Alexandrou:2002sn,deForcrand:2005vv}.
In particular, it is possible to directly compute the gluon field
distribution \cite{Ichie:2002mi,Okiharu:2003vt,Bissey:2005sk} using
lattice QCD techniques similar to those pioneered in mesonic
static-quark systems \cite{Sommer:1987uz,Bali:1994de,Haymaker:1994fm}.

Similar to Okiharu and Woloshyn \cite{Okiharu:2003vt}, our first
interest is to test the static-quark source-shape dependence of the
observed flux distribution, as represented by correlations between the
quark positions and the action or topological charge density of the
gauge fields.  To this end, we choose three different ways of
connecting the gauge link paths required to create gauge-invariant
Wilson loops as in our preliminary study \cite{Bissey:2005sk}.  In the
first case, quarks are connected along a T-shape path, while in the
second case an L-shape is considered.  Finally symmetric link paths
approximating a Y-shape such that the quark positions approximate the
vertices of an equilateral triangle are considered.  The latter is
particularly interesting as the probability of observing a
$\Delta$-shape flux-tube (empty triangle) distribution is maximised in
this equidistant case.  The most recent lattice QCD studies are in
favor of the Y ansatz when it comes to the confinement potential
\cite{Takahashi:2002bw} at long distances
\cite{Alexandrou:2001ip,Alexandrou:2002sn,deForcrand:2005vv} and flux
tube formation \cite{Ichie:2002mi}.  We also note that a theoretical
calculation using center-vortex QCD with gauge group $SU(3)$ also
favors the Y ansatz \cite{Cornwall:2006}.  In order to better compare
T and Y shapes we study configurations in which the quarks are in the
same location and only the connection point changes. Such a direct
comparison is not possible with the L-shape.

Because the signal decreases in an exponential fashion with the size
of the loop, it is essential to use a method to enhance overlap with
the ground state of interest.  We use strict three-dimensional APE
smearing \cite{APE87} to enhance the overlap of our spatial-link paths
with the ground-state static-quark baryon potential.  Time-oriented
links remain untouched to preserve the correct static quark potential
at all separations.  We present results for two levels of source
smearing including a minimal smearing of 10 APE steps and an optimal
smearing of 30 APE steps.  The comparison of two levels of smearing
reveals interesting insight into the manner in which the flux tubes
evolve to become the ground state.

Complete details of the construction of Wilson loops for the
determination of the static quark potential and associated gluon-field
distributions are presented in Sec.~\ref{wilsonLoops}.
Sec.~\ref{correlator} presents the three-point correlation function
calculated to reveal the response of the vacuum to the presence of
quarks.  Of particular importance is the insensitivity to the choice
of metric such that the observation of vacuum field suppression is
unambiguous. 

Excellent signal to noise is achieved via a high-statistics approach
based on translational symmetry of the four-dimensional lattice volume
and rotational symmetries of the lattice as described in detail in
Sec.~\ref{stats}.   This approach contrasts 
with previous investigations, which used 
gauge fixing followed by projection to smooth the links and resolve a
signal in the flux distribution \cite{Ichie:2002mi,Okiharu:2003vt}.

Sec.~\ref{results10} presents the simulation results for sources
constructed from 10-sweep smeared links.  We find substantial
correlation between the source and the observed field distribution.
While these sources provide information vital to linking the flux-tube
distribution to the static quark potential, they are not adequate to
isolate the ground-state potential at large quark separations.
Therefore in Sec.~\ref{results30} we complement these results with
sources constructed from optimal 30-sweep smeared links.  Here clear
deformation of the T-shape source to a Y-shape configuration is
illustrated.  A quantitative analysis comparing the static quark
potentials is used to identify the ground state and in
Sec.~\ref{tubeProps} the position, radius and expulsion extent of
the flux tubes and the node connecting them are determined.
A summary of our findings is provided in Sec.~\ref{conclusions}.

\section{WILSON LOOPS}
\label{wilsonLoops}

\begin{figure}
\centering\includegraphics[clip=true]{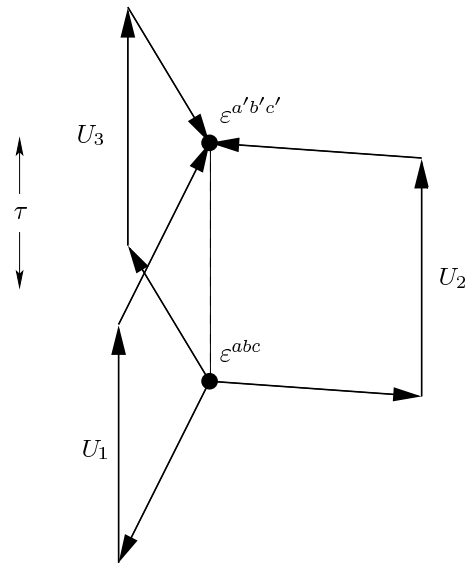}
\caption{Gauge-link paths or ``staples,'' $U_1$, $U_2$ and $U_3$,
  forming a three-quark Wilson loop with the quarks located at $\vec
  r_1$, $\vec r_2$ and $\vec r_3$.  $\varepsilon^{abc}$ and
  $\varepsilon^{a'b'c'}$ denote colour anti-symmetrisation at the source
  and sink respectively, while $\tau$ indicates evolution of the
  three-quark system in Euclidean time.}
\label{staple}
\end{figure}

To study the flux distribution in baryons on the lattice, one begins
with the standard approach of connecting static quark propagators by
spatial-link paths in a gauge invariant manner.  APE-smeared
spatial-link paths propagate the quarks from a common origin to their
spatial positions as illustrated in Fig.~\ref{staple}.  The static
quark propagators are constructed from time directed link products at
fixed spatial coordinate, $\prod_i U_t(\vec x, t_i)$, using the
untouched ``thin'' links of the gauge configuration.

The smearing procedure replaces a spatial link, $U_{\mu}(x)$, with a
sum of $1-\alpha$ times the original link plus $\alpha/4$ times its
four spatially oriented staples, followed by projection back to
$SU(3)$.  We select the special-unitary matrix $U_{\mu}^{\rm FL}$
which maximises $ {\cal R}e \, {\rm{tr}}(U_{\mu}^{\rm FL}\,
U_{\mu}'^{\dagger})$, where $U_{\mu}'$ is the smeared link, by
iterating over the three diagonal $SU(2)$ subgroups of $SU(3)$
repeatedly.  In practice, eight iterations over the three subgroups
optimizes the projected link to about 1 part in $10^8$.

We initially repeated the combined procedure of smearing and
projection $10$ times, with $\alpha = 0.7$.  However, it will become
clear that $10$ times is not enough to isolate the ground state of the
three-quark static-quark potential at large quark separations.  We
have found that repeating the smearing-projection procedure $30$ times
is effective in isolating the ground-state.  In this case the
string-tension of the three-quark potential matches that of the
quark-antiquark potential.  We will present for comparison, results
for both $10$ and $30$ steps of APE-source smearing.

Untouched links in the time direction propagate the spatially
separated quarks through Euclidean time.  In principle, the ground
state is isolated after sufficient time evolution.  Finally
smeared-link spatial paths propagate the quarks back to the common
spatial origin.

The three-quark Wilson loop is defined as:
\begin{equation}
W_{3Q}=\frac{1}{3!}\varepsilon^{abc}\varepsilon^{a'b'c'} \, U_1^{aa'} \,
U_2^{bb'} \, U_3^{cc'},
\end{equation}  
where $U_j$ is a staple made of path-ordered link variables
\begin{equation}
U_j \equiv P \exp \left( ig \int_{\Gamma_j} dx_{\mu} \, A^{\mu}(x)
\right) \, ,
\end{equation}
and $\Gamma_j$ is the path along a given staple as shown in 
Fig.~\ref{staple}.

\begin{figure}
\centering\includegraphics[height=2cm,clip=true]{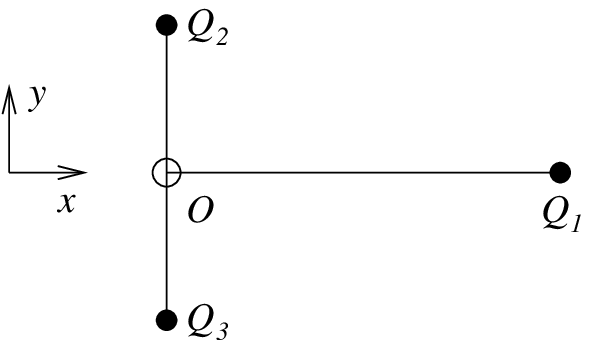}
\caption{Projection of the T-shape path on the $x$-$y$ plane.}
\label{tpath}
\centering\includegraphics[height=2cm,clip=true]{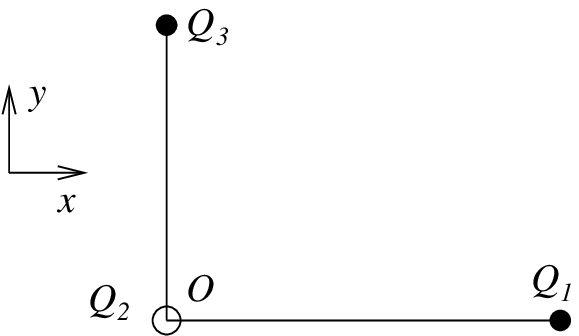}
\caption{Projection of the L-shape path on the $x$-$y$ plane.}
\label{lpath}
\centering\includegraphics[height=2cm,clip=true]{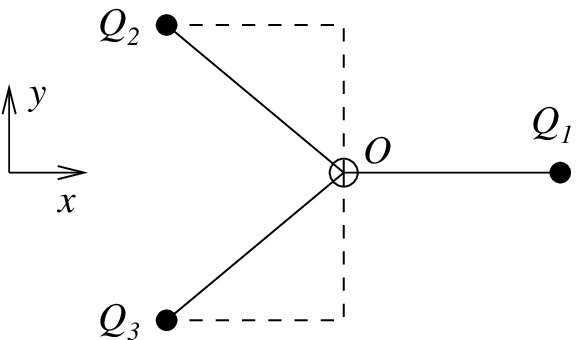}
\caption{Projection of the Y-shape path on the $x$-$y$ plane.}
\label{ypath}
\end{figure}

In this study, we consider several two-dimensional spatial-link
paths. Figures 
\ref{tpath} and \ref{lpath} give the projection of T and L shape paths
in the $x$-$y$ plane. Figure \ref{ypath} shows the paths considered in
constructing the Y-shape.  In all cases the three quarks are created
at the origin, $O$ (white bubble), then are propagated to the positions
$Q_1$, $Q_2$ or $Q_3$ (black circle) before being propagated through
time and finally back to a sink at the same spatial location as the
source ($O$).

\begin{figure}
\centering\includegraphics[height=1.3cm,clip=true]{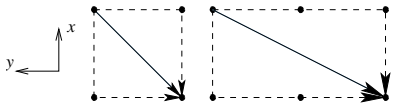}
\caption{The diagonal path (solid line) is taken as the average of
the two possible paths going around the box enclosing the end points.}
\label{yboxes}
\end{figure}

For the Y-shape we create elementary diagonal ``links'' in the form of
boxes as shown in Fig. \ref{yboxes}.  The $1 \times 1$ and $1 \times
2$ boxes are the average of the two path-ordered link variables going
from one corner to the diagonally opposite one. Taking both of these
paths better maintains the symmetry of the ground state potential and
therefore provides improved overlap with the ground state.  We also
consider $2 \times 3$ boxes which are the averages of the possible
paths connecting two opposite corners using $1\times 1$ and $1\times
2$ boxes.  We may create further link paths in the future for use in
bigger loops if it becomes necessary.  Hence a diagonal staple is, in
fact, an average of several ``squared path'' staples connecting the
same end points.

The quark coordinates considered for Y and T shapes paths are
summarised in Table~\ref{tab:coord}.  The Fermat point -- the point
within an acute triangle that minimizes the sum of the distances to
the vertices -- is also indicated. We note that the origin $(0,0)$
is not at the centre of these coordinates.  Rather the coordinates are
selected to place the quarks at approximately equal distances from
each other. As one can see, the Fermat point, $F$, (also called the
Steiner point in this particular case) is very close to the origin in
most cases. In the table we also indicate the average inter-quark
distance $\langle d_{qq} \rangle$ and the average distance to the
Fermat point $\langle r_s \rangle$.

Note that in comparison to the measures considered in
Ref.~\cite{Alexandrou:2002sn}, we have $L_Y = 3\, \langle r_s \rangle$
and $L_\Delta =3\, \langle d_{qq} \rangle$.  If the quark positions were
on the vertices of an equilateral triangle, $L_Y$ and $L_\Delta$ would
be linked in a simple linear fashion.  In the quark configurations of
Table~\ref{tab:coord}, the quantity $\langle r_s \rangle$ is always
within one part in a thousand of the radius of the circle
circumscribing the triangle, therefore $\langle r_s \rangle$ is a good
measure of the size of the baryon.

For the L-shape we have studied a small sample of 11 configurations,
with $10$-step smearing.  The following coordinates for the quarks are
considered: $Q_1 = (0,\ell)$, $Q_2 = (0,0)$ and $Q_3 = (\ell,0)$ where
$\ell$ goes from $1$ to $11$.  Because all L-shape triangles are
simply a scaled version of a unique triangle, all distances in this
shape are a given constant times the scale factor (here ``$\ell$'') as
in an equilateral triangle.  In lattice units, we have $F
=(\frac{3-\sqrt{3}}{6}\ell,\frac{3-\sqrt{3}}{6}\ell)$, $\langle r_s
\rangle = \frac{1+\sqrt{3}}{3\sqrt{2}}\ell$, $\langle d_{qq} \rangle =
\frac{2+\sqrt{2}}{3}\ell$ and the radius of the circumscribed circle
is $\frac{\ell}{\sqrt{2}}$. The L-shapes are far further away from the
ideal equilateral triangle than the T and Y shapes and $\langle r_s
\rangle$ is not a good measure of the size of the baryon.

\begin{table}
\caption{ $(x,y)$ coordinates for the three quarks considered for Y-
  and T-shape paths.  We also show the coordinates of the Fermat
  point, $F$, and the average distance of the quarks
  from this point $\langle r_s \rangle $.  Also shown is the average
  inter-quark distance $\langle d_{qq} \rangle$.}
\label{tab:coord}
\begin{ruledtabular}
\begin{tabular}{lcccccc}
&\multicolumn{4}{c}{$(x,y)$ Coordinates (lattice units)} 
         &\multicolumn{2}{c}{Distance (fm)} \\
\# &$Q_1$  &$Q_2$  &$Q_3$  & $F$ &$\quad\langle r_s \rangle\quad $ &
$\langle d_{qq}\rangle $ \\
\noalign{\smallskip}
\hline
\noalign{\smallskip}
1 &$( 1, 0)$ &$(-1, 1)$ &$(-1,-1)$ & $(-0.42, 0)$ &  0.15  & 0.27 \\
2 &$( 2, 0)$ &$(-1, 2)$ &$(-1,-2)$ & $( 0.15, 0)$ &  0.27  & 0.46 \\
3 &$( 3, 0)$ &$(-1, 2)$ &$(-1,-2)$ & $( 0.15, 0)$ &  0.31  & 0.53 \\
4 &$( 3, 0)$ &$(-2, 3)$ &$(-2,-3)$ & $(-0.27, 0)$ &  0.42  & 0.72 \\
5 &$( 4, 0)$ &$(-3, 4)$ &$(-3,-4)$ & $(-0.69, 0)$ &  0.57  & 0.99 \\
6 &$( 5, 0)$ &$(-4, 5)$ &$(-4,-5)$ & $(-1.11, 0)$ &  0.72  & 1.25 \\
7 &$( 7, 0)$ &$(-4, 6)$ &$(-4,-6)$ & $(-0.54, 0)$ &  0.88  & 1.52 \\
8 &$( 8, 0)$ &$(-4, 7)$ &$(-4,-7)$ & $( 0.04, 0)$ &  0.99  & 1.71 \\
\end{tabular}
\end{ruledtabular}
\end{table}

\section{GLUON FIELD CORRELATION}
\label{correlator}

In this investigation we characterise the gluon-field fluctuations by
the gauge-invariant action density $S(\vec y, t)$ observed at spatial
coordinate $\vec y$ and Euclidean time $t$ measured relative to the
origin of the three-quark Wilson loop.  We calculate the action
density using the highly-improved ${\cal O}(a^4)$ three-loop improved
lattice field-strength tensor \cite{Bilson-Thompson:2002jk} on
four-sweep APE-smeared gauge links.  While the use of this
highly-improved action will suppress correlations immediately next to
the quark positions, it will assist tremendously in revealing the
flux-tube correlations which are the central focus of this
investigation.

Defining the quark positions as $\vec r_1$, $\vec r_2$
and $\vec r_3$ relative to the origin of the three-quark Wilson loop,
and denoting the Euclidean time extent of the loop by $\tau$, we evaluate
the following correlation function
\begin{eqnarray}
\lefteqn{C(\vec y; \vec r_1, \vec r_2, \vec r_3; \tau) =} \nonumber \\
&&\quad \frac{
\bigl\langle W_{3Q}(\vec r_1, \vec r_2, \vec r_3; \tau) \,
             S(\vec y, \tau/2) \bigr\rangle }
{
\bigl\langle  W_{3Q}(\vec r_1, \vec r_2, \vec r_3; \tau) \bigr\rangle \,
\bigl\langle S(\vec y, \tau/2) \bigr\rangle
}
  \, ,
\label{correl}
\end{eqnarray} 
where $\langle \cdots \rangle$ denotes averaging over configurations
and lattice symmetries as described below.
This formula correlates the quark positions via the three-quark Wilson
loop with the gauge-field action in a gauge invariant manner.  For
fixed quark positions and Euclidean time, $C$ is a scalar field in
three dimensions.  For values of $\vec y$ well away from the quark
positions $\vec r_i$, there are no correlations and $C \to 1$.

This measure has the advantage of being positive definite, eliminating
any sign ambiguity on whether vacuum field fluctuations are enhanced
or suppressed in the presence of static quarks.  We find that $C$ is
generally less than 1, signaling the expulsion of vacuum fluctuations
from the interior of heavy-quark hadrons.

\section{STATISTICS}
\label{stats}

In this investigation we consider 200 quenched QCD gauge-field
configurations created with the ${\cal O}(a^2)$-mean-field improved
Luscher-Weisz plaquette plus rectangle gauge action
\cite{Luscher:1984xn} on $16^3\times 32$ lattices at $\beta =4.60$.
The long dimension is taken as being the $x$ direction making
the spatial volume $16^2\times 32$.
Using a physical string tension of $\sigma = (0.440\ \text{GeV})^2 =
0.981\ \text{GeV}/\text{fm}$, the $q\bar{q}$ potential sets the lattice
spacing to $a=0.123(2)$fm.  In our previous study \cite{Bissey:2005sk}
we used $12^3\times 24$ lattices with the same action and parameter
$\beta$.

To improve the statistics of the simulation we use various symmetries
of the lattice.  First, we make use of translational invariance by
computing the correlation on every node of the lattice, averaging the
results over the four-volume.

To further improve the statistics, we use reflection symmetries.
Through reflection on the plane $x=0$ we can double the number of T
and Y-shaped Wilson loops. By using reflections on both the plane
$x=0$ and $y=0$ we can quadruple the number of L-shaped loops.

We finally use $90^\circ$ rotational symmetry about the $x$-axis to
double the number of Wilson loops. This means we are using both the
$x$-$y$ and $x$-$z$ planes as the planes containing the quarks.

In summary, the Y and T shape Wilson loop results are the average of
$16^3\times 32 \times 2 \times 2 \times 200 = 104,857,600$ Wilson loop
calculations.

\section{SIMULATION RESULTS WITH 10 SWEEPS SMEARING}
\label{results10}

\subsection{Qualitative results}

\begin{figure}
\centering\includegraphics[height=6.2cm,clip=true]{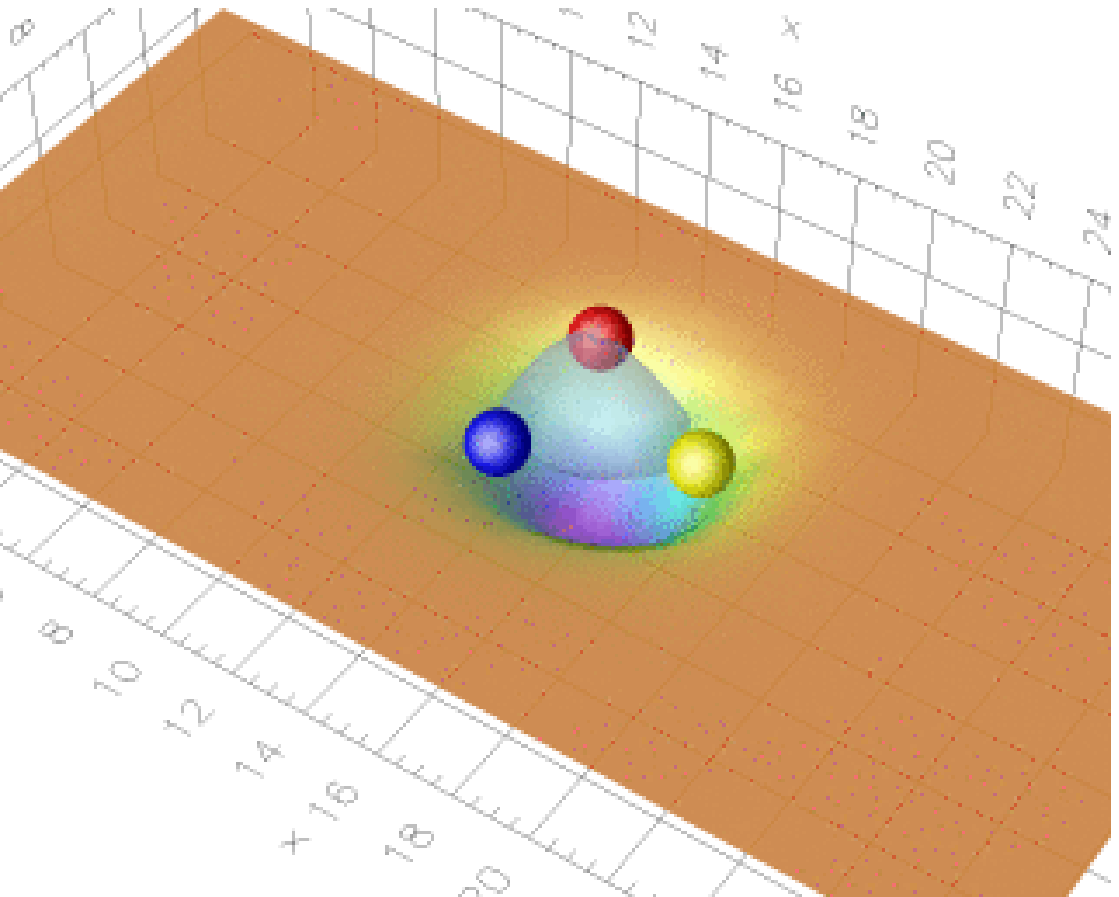}
\caption{Expulsion of gluon-field fluctuations from the region of
  static quark sources illustrated by the spheres.  An isosurface of
  $C(\vec{y})$ is illustrated by the translucent surface.  A
  surface plot (or rubber sheet) describes the values of $C(\vec{y})$
  for $\vec y$ in the quark plane, $(y_1, y_2, 0)$.  Results are for a
  10-sweep smeared T-shape source with quark positions as in the third
  configuration of Table \ref{tab:coord}. The maximum expulsion is 9.7\%
  and the isosurface is set to 4.9\%.}
\label{Ttube3}
\centering\includegraphics[height=6.2cm,clip=true]{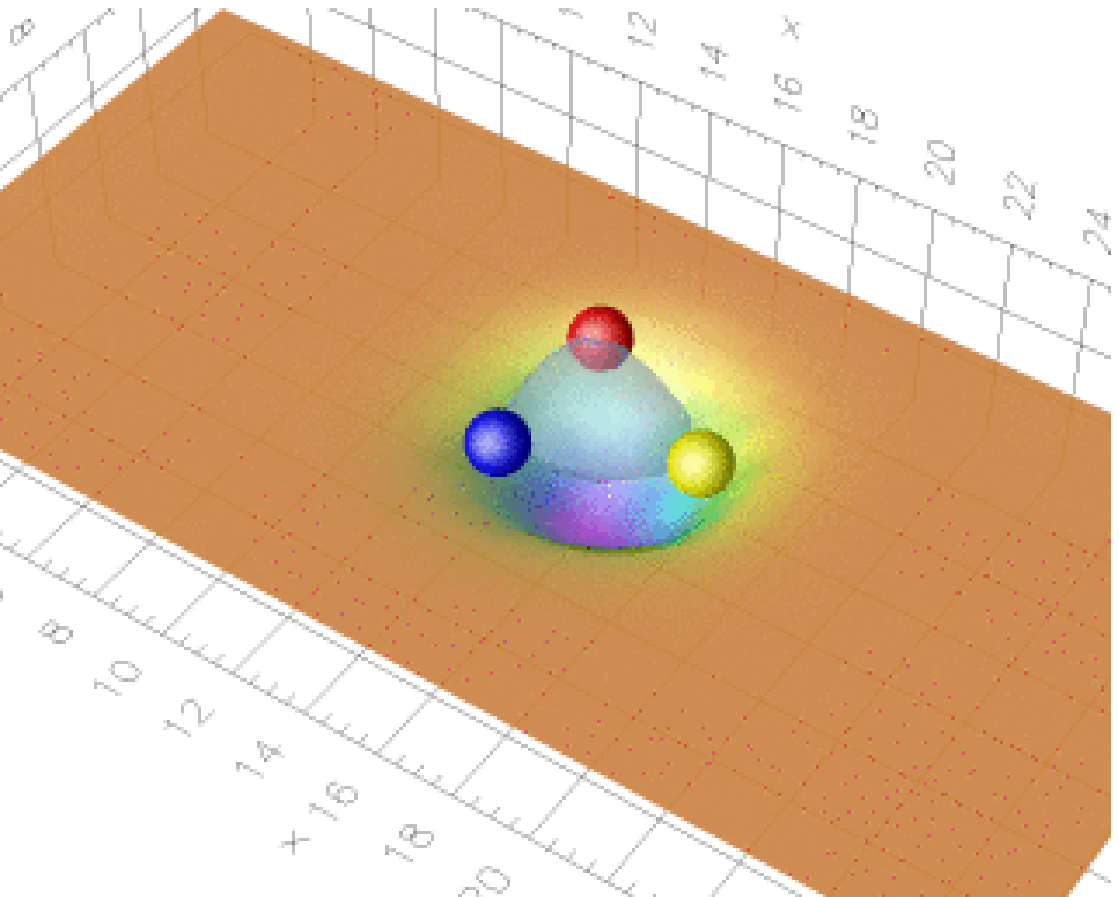}
\caption{Isosurface and surface plot of $C(\vec{y})$ for a 10-sweep
   smeared Y-shape source with quark positions as in the third
   configuration of Table \ref{tab:coord}. The maximum expulsion is
   9.8\% and the isosurface is set to 4.9\%. Further details are described
   in the caption of Fig.~\protect\ref{Ttube3}.}
\label{Ytube3}
\centering\includegraphics[height=6.2cm,clip=true]{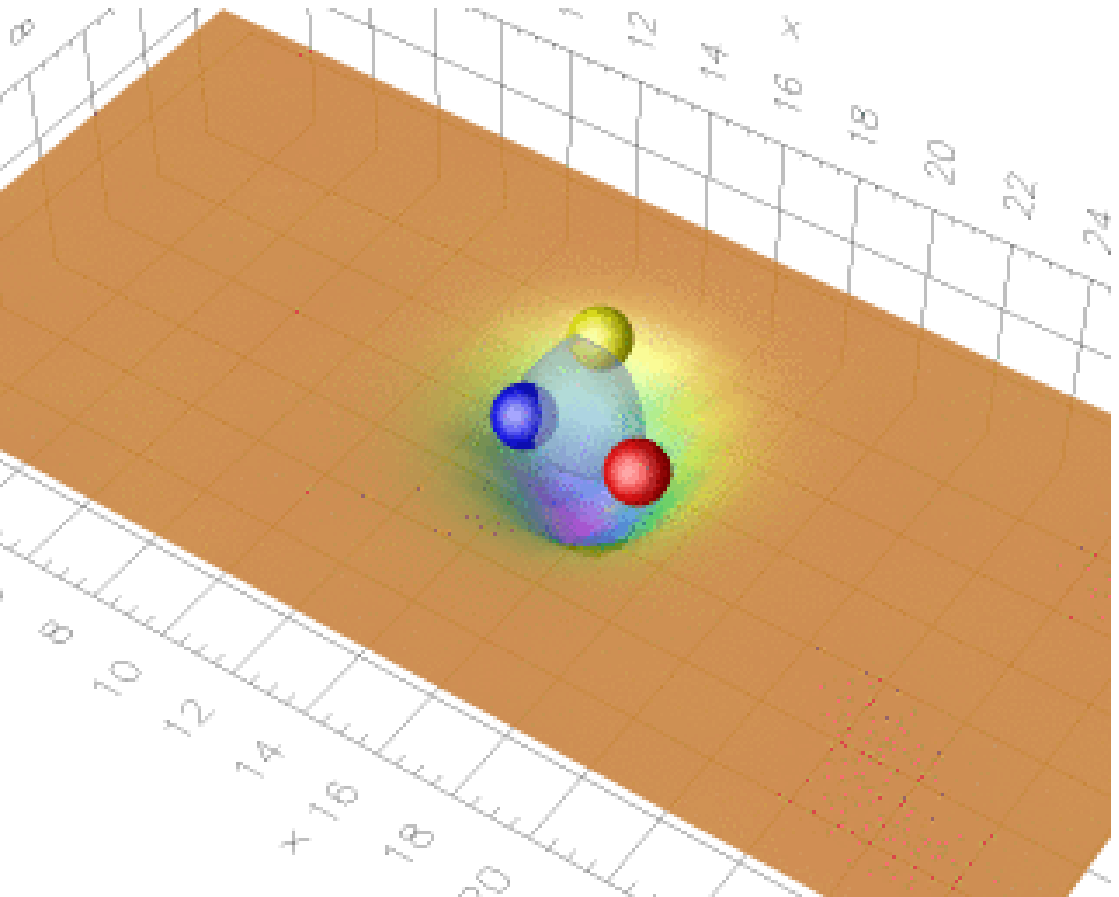}
\caption{Isosurface and surface plot of $C(\vec{y})$ for a 10-sweep
  smeared L-shape source with quark separations of $\ell=3$. 
  The maximum expulsion is 9.5\% and the isosurface is set to 5.1\%.
  Further details are described in the caption of Fig.~\protect\ref{Ttube3}.}
\label{Ltube3}
\end{figure} 

\begin{figure}
\centering\includegraphics[height=6.2cm,clip=true]{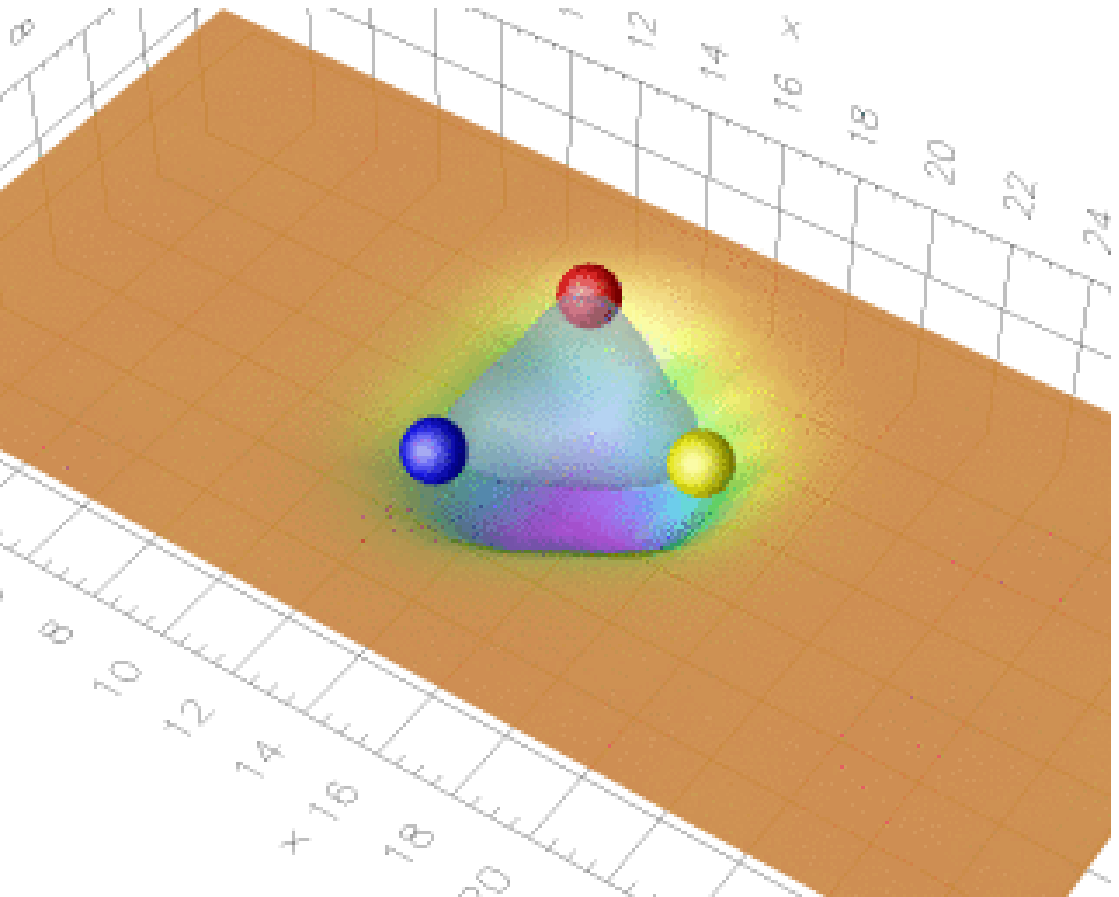}
\caption{Isosurface and surface plot of $C(\vec{y})$ for a 10-sweep
   smeared Y-shape source with quark positions as in the fourth
   configuration of Table \ref{tab:coord}. The maximum expulsion is
   8.9\% and the isosurface is set to 4.5\%. Further details are described
   in the caption of Fig.~\protect\ref{Ttube3}.}
\label{Ytube4}
\end{figure}

\begin{figure}
\centering\includegraphics[height=6.2cm,clip=true]{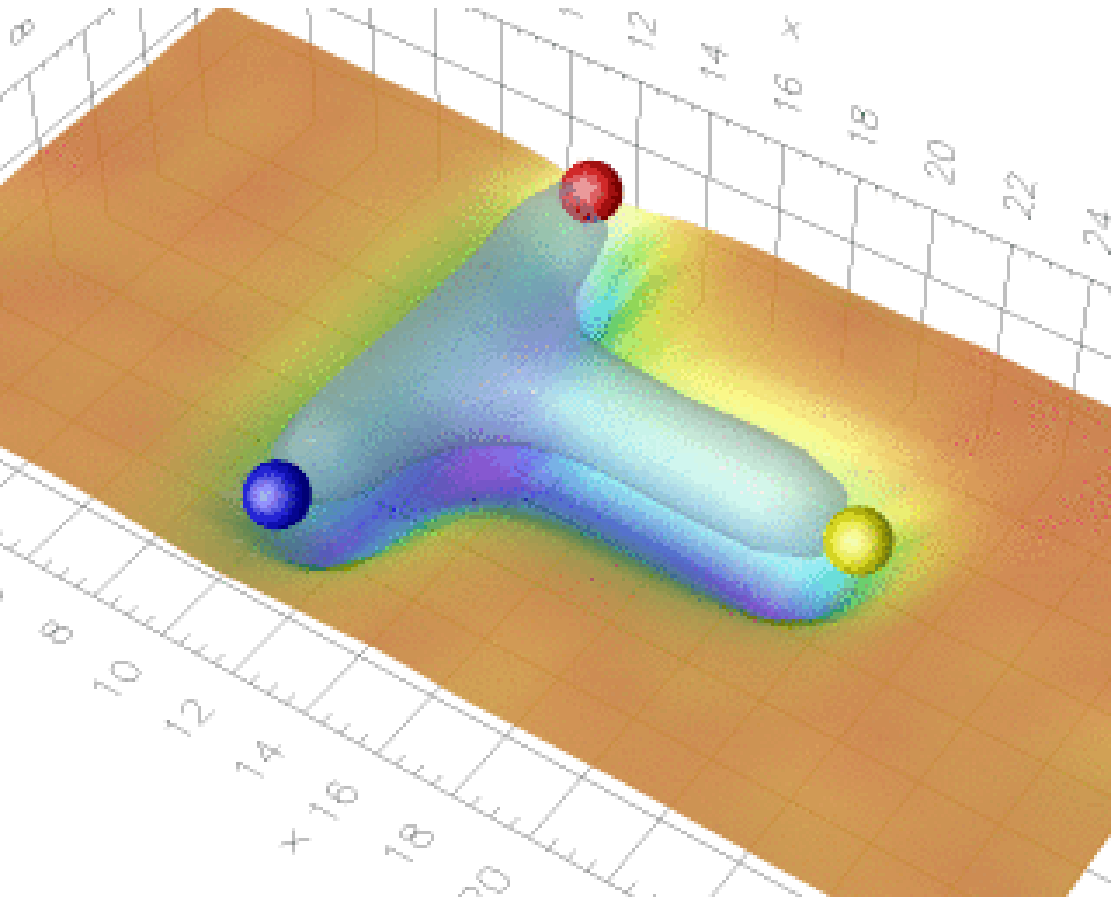}
\caption{Isosurface and surface plot of $C(\vec{y})$ for a 10-sweep
   smeared T-shape source with quark positions as in the seventh
   configuration of Table \ref{tab:coord}. The maximum expulsion is
   8.3\% and the isosurface is set to 4.4\%. Further details are described
   in the caption of Fig.~\protect\ref{Ttube3}.}
\label{Ttube7}
\centering\includegraphics[height=6.2cm,clip=true]{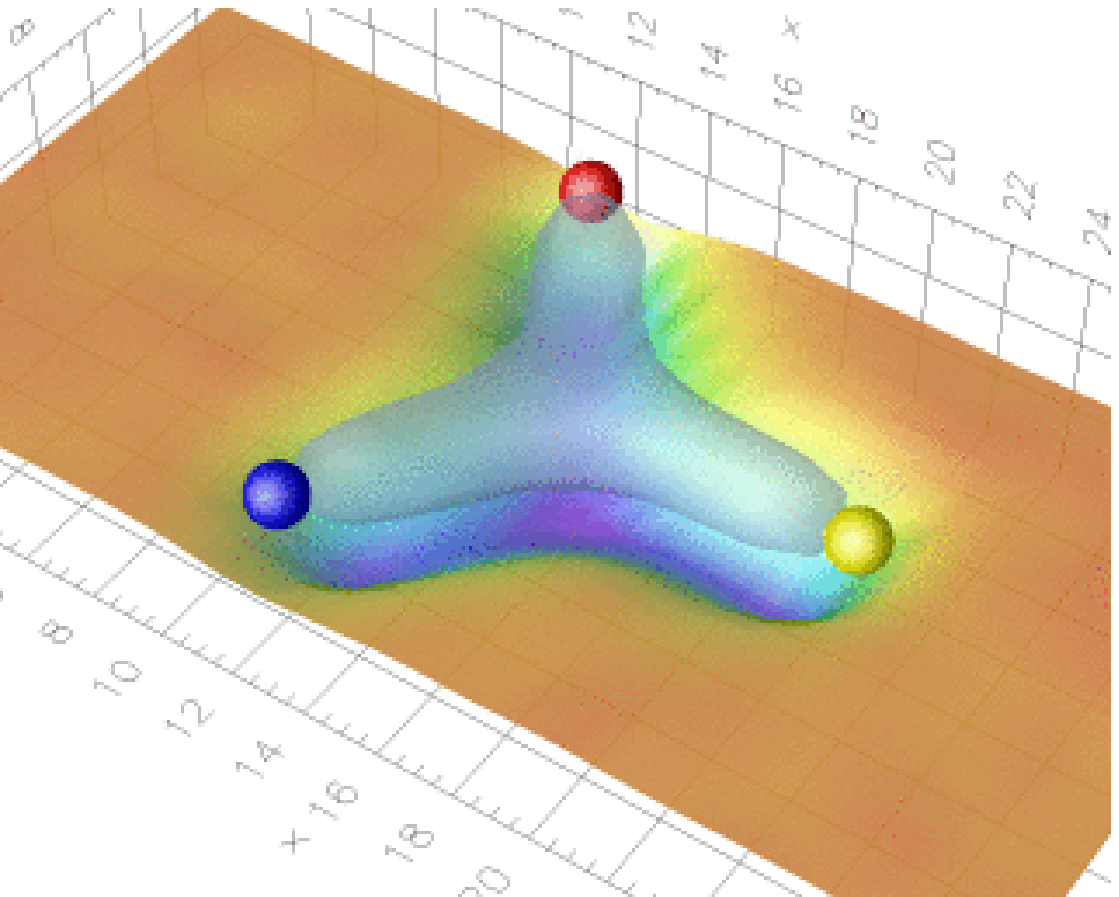}
\caption{Isosurface and surface plot of $C(\vec{y})$ for a 10-sweep
   smeared Y-shape source with quark positions as in the seventh
   configuration of Table \ref{tab:coord}. The maximum expulsion is 8.3\%
   and the isosurface is set to 4.4\%. Further details are described
   in the caption of Fig.~\protect\ref{Ttube3}.}
\label{Ytube7}
\centering\includegraphics[height=6.2cm,clip=true]{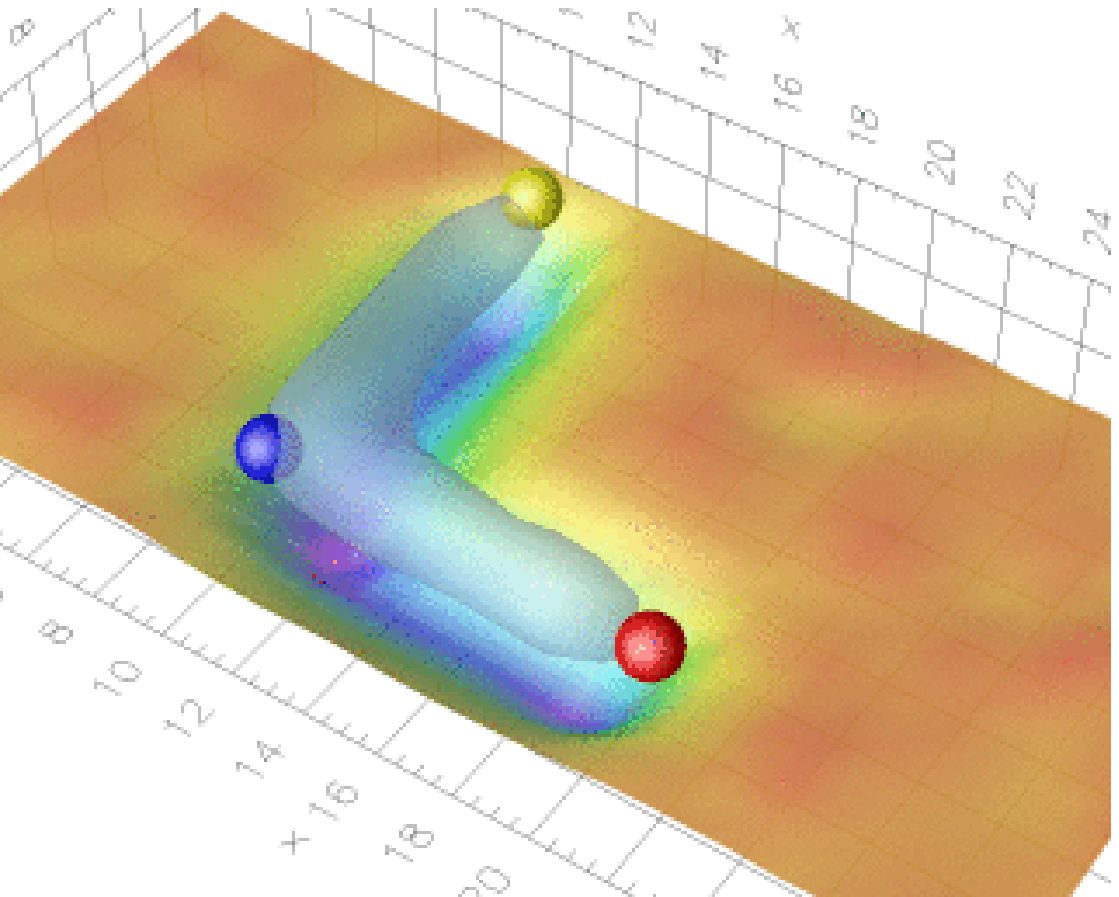}
\caption{Isosurface and surface plot of $C(\vec{y})$ for a 10-sweep
  smeared L-shape source with quark separations of $\ell=10$. 
  The maximum expulsion is 8.8\% and the isosurface is set to 4.4\%.
  Further details are described in the caption of Fig.~\protect\ref{Ttube3}.}
\label{Ltube10}
\end{figure}

\begin{figure}
\centering\includegraphics[width=9.0cm,bb= 0 52 320 240,clip=true]{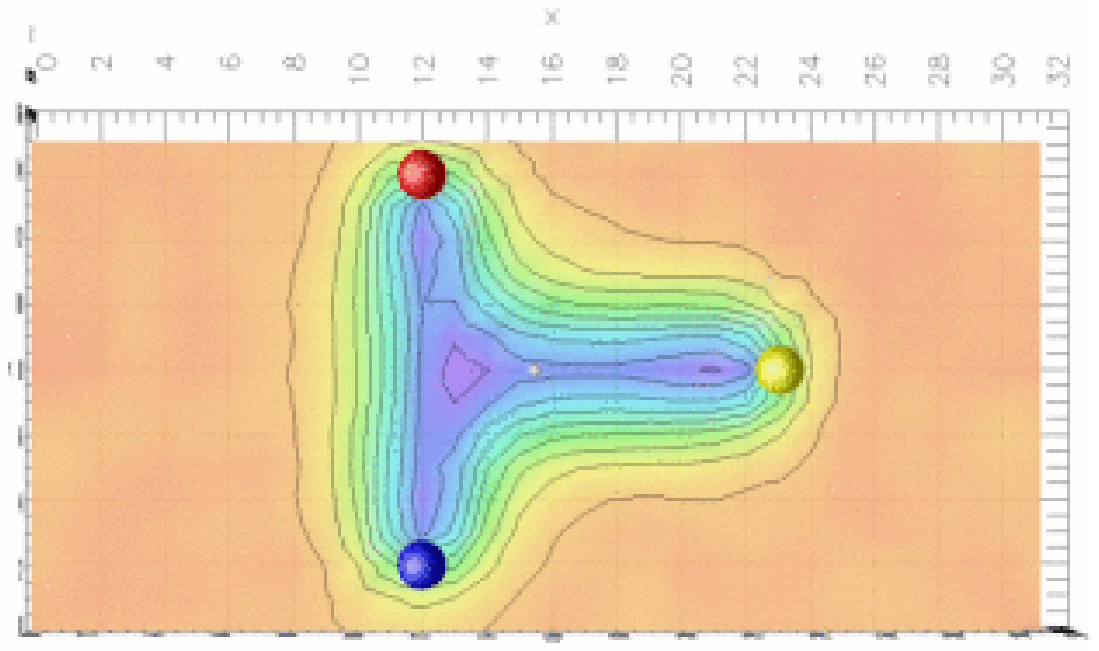}
\caption{Contour plot of $C(\vec{y})$ for $\vec y$ in the quark plane,
  $(y_1, y_2, 0)$, for a 10-sweep smeared T-shape source with quark
  positions as described in the seventh configuration of Table
  \ref{tab:coord}.  The white diamond shows the position of the Fermat
  point.}
\label{Tshape7}
\centering\includegraphics[width=8.5cm,bb= 0 52 320 240,clip=true]{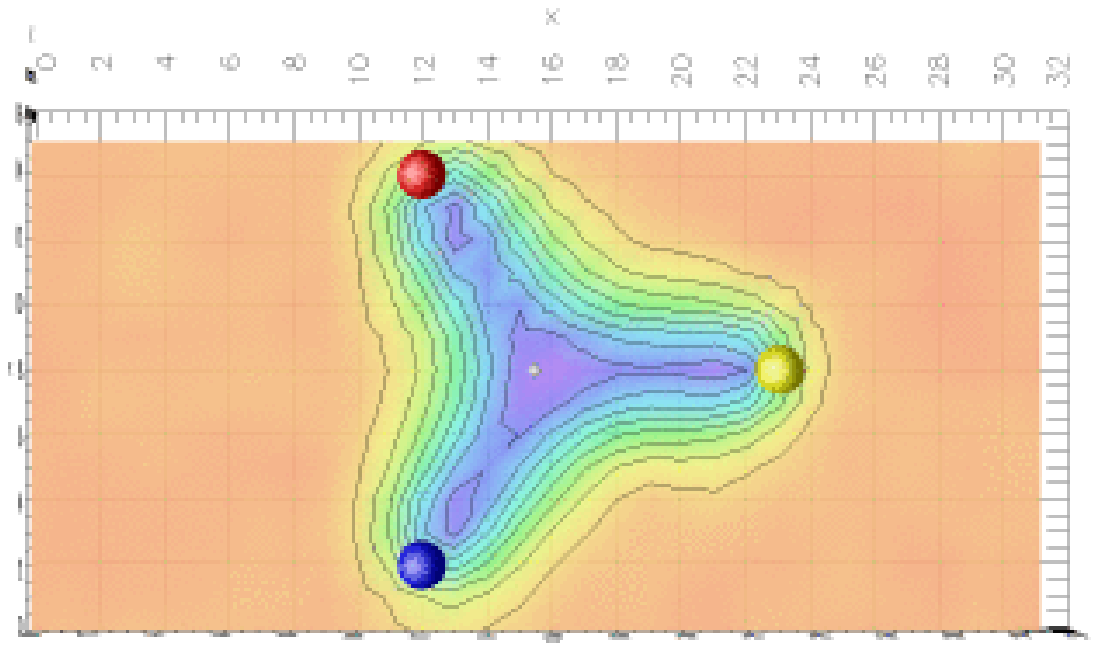}
\caption{Contour plot of $C(\vec{y})$ for $\vec y$ in the quark plane,
  $(y_1, y_2, 0)$, for a 10-sweep smeared Y-shape source with quark
  positions as described in the seventh configuration of Table
  \ref{tab:coord}.  The white diamond shows the position of the Fermat
  point.}
\label{Yshape7}
\centering\includegraphics[width=8.5cm,bb= 0 52 320 240,clip=true]{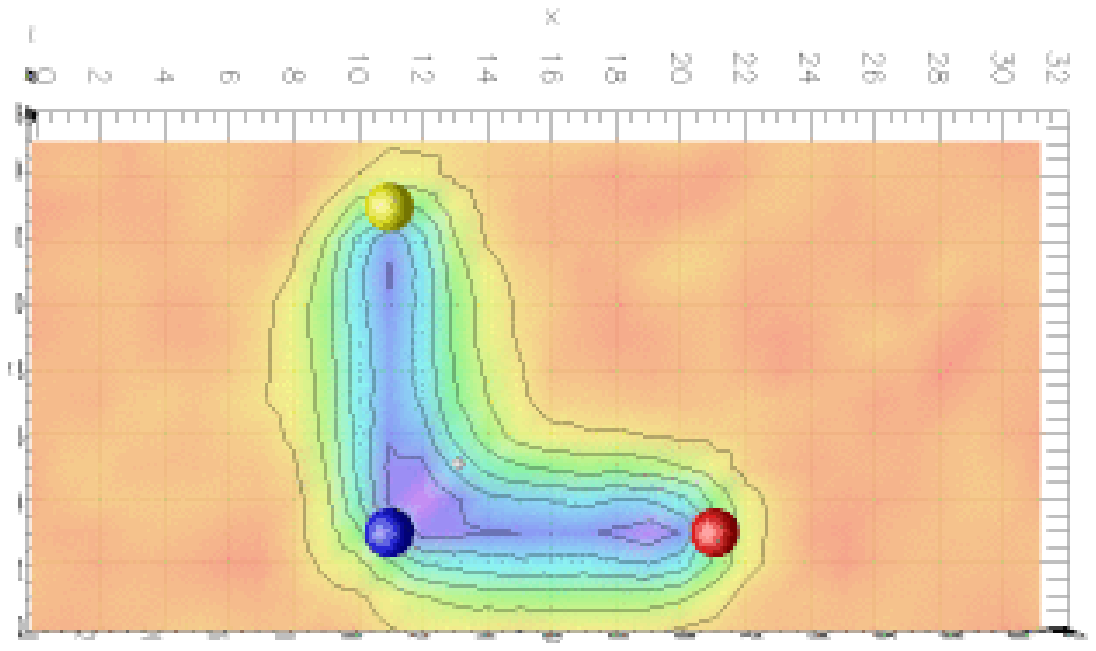}
\caption{Contour plot of $C(\vec{y})$ for $\vec y$ in the quark plane,
  $(y_1, y_2, 0)$, for a 10-sweep smeared L-shape source with quark
  separations of $\ell=10$. The white diamond shows the position of
  the Fermat point.}
\label{Lshape10}
\end{figure}

We begin by examining the correlation function $C(\vec y; \vec r_1,
\vec r_2, \vec r_3; \tau)$ for 10-sweep smeared T-, Y- and L-shape
sources at $\tau = 2$.  Larger time evolutions display the same
features and are discussed further below.

At small quark separations, both the T- and Y-shape sources reveal a
similar response of the vacuum to the quark positions.  As seen in
Figs.~\ref{Ttube3} and \ref{Ytube3}, isosurfaces of equal $C(\vec{y})$
form a bubble for quark positions in the third configuration of Table
\ref{tab:coord}.  Similar results are seen for smaller separations.

The isosurface for lower levels of field expulsion ({\it i.e.} further away
from the quark positions) displays an approximately spherical shape as
suggested by the surface plot for values of $\vec y$ just outside the
quark positions.

In these figures we also show the value of $C(\vec{y})$ for $\vec y$
in the quark plane, $(y_1, y_2, 0)$, on a surface plot (or rubber
sheet).  We clearly see that $C(\vec{y})$ is less than $1$ in the
bubble, indicating an expulsion or suppression of vacuum gluon-field
fluctuations.

These features are also seen to some extent for the L shape up until
$\ell=3$ as shown in Fig. \ref{Ltube3}.  These results correspond to a
radius of about 0.3 fm or under.

At these small separations, one has the luxury of exploring the
Euclidean time evolution of $C(\vec y; \vec r_1, \vec r_2, \vec r_3;
\tau)$.  While the depth of the vacuum field expulsion increases in
going from $\tau = 2$ to 4, it saturates in going from 4 to 6.
In all cases however, there is no discernible change in the shape of
the expulsion.  Normalized contour plots of $C(\vec y; \tau)$ are
independent of $\tau$ to an excellent approximation.

With the fourth set of quark coordinates the bubble loses more of its
symmetry and a filled triangular shape emerges as in
Fig.~\ref{Ytube4}, which corresponds to a radius of about 0.4 fm. This
transition occurs at a radius of 0.35 fm for the L shape.  However,
this shape is much less symmetric to start with, as one of the
inter-quark distances is 41\% larger than the other two.

It is interesting to note the expulsion indicated by the suppression
of $C(\vec{y})$ is largest at the centre of the quark configuration
for all four quark-source shapes illustrated in Figs.~\ref{Ttube3}
through \ref{Ytube4}.  As such, there is no evidence of a
$\Delta$-shape flux-tube (empty triangle) distribution forming at
small distances.  Rather, the vacuum expulsion observed in
%However, the vacuum expulsion itself as seen on 
Fig.~\ref{Ytube4} is shaped as a solid or filled triangle as opposed
to tubes.  Here the inter-quark distance is more than 0.7 fm, which is
outside the region where the potential is dominated by one-gluon
exchange.

For radii larger than 0.5 fm we clearly see the formation of flux
tubes as shown in Figs.~\ref{Ttube7} to \ref{Ltube10}.  Here, a robust
dependence of the observed flux-tube shape on the shape of the
source is revealed.  However, such a correlation is not surprising as
these  figures correspond to $\tau = 2$.  In principle, with
sufficient time evolution these images will relax into the same
ground-state configuration.  In practice this is not possible as the
signal is lost rapidly.  Instead, one turns one's attention to
improving the source to best isolate the ground state, as we shall do
in the following sections.

Nevertheless, some evolution of the flux tubes away from the source
shape towards the Y-shape is already apparent.
For example the flux-tube node of the T-shape source has moved by at
least one lattice unit towards the Fermat point as illustrated in the
contour plot of Fig.~\ref{Tshape7}.  Some curvature in the ``top'' of
the T (left-most in the figure) is also apparent.
In contrast, the observed Y shape of Fig.~\ref{Yshape7} is quite close
to the source shape and the observed node of the flux tubes, which is
very close to the origin, appears to have moved towards the Fermat
point.

For the L shape results of Fig.~\ref{Lshape10} the issue is more
subtle.  While the node appears to have moved in the direction of the
Fermat point to the next grid intersection, this observation might
also be due to the suppression of the correlation near the quark
positions associated with our use of a highly improved action.

While, Figs. \ref{Ttube3} through \ref{Lshape10} show the action-based
correlation function results, we have also examined correlations of
the Wilson loops with the topological charge, the squared colour
magnetic field and squared colour electric fields, all of which are
gauge invariant.  The results are qualitatively similar and in some
cases indistinguishable.

\subsection{Quantitative results}

We now move to a quantitative study of these results exploiting the
knowledge of the node position from our qualitative analysis.  We
begin by determining the effective three-quark potential for the
various quark positions, source shapes and Euclidean time evolutions.
The vacuum expectation value for $W_{3Q}$ is
\begin{equation}
\langle W_{3Q} (\tau) \rangle = \sum_{n=0}^\infty C_n \exp(-a\, V_n \,
\tau),
\label{expec}
\end{equation}
where $V_n$ is the potential energy of the $n$-th excited state and
$C_n$ describes the overlap of the source with the $n$-th state.  The
effective potential is extracted from the Wilson loop via the standard
ratio
\begin{equation}
a\, V(\vec{r},\tau)= \ln\left(
\frac{W_{3Q}(\vec{r},\tau)}{W_{3Q}(\vec{r},\tau+1)}\right).
\label{pot}
\end{equation}
If the ground state is indeed dominant, plotting $V$ as a function of
$\tau$ will show a plateau and any curvature can be associated with
excited state contributions.  Statistical uncertainties are estimated
via the jackknife method \cite{Montvay:1994bk}.

\begin{figure}[tbhp]
\centering\includegraphics[width=8.4cm,clip=true]{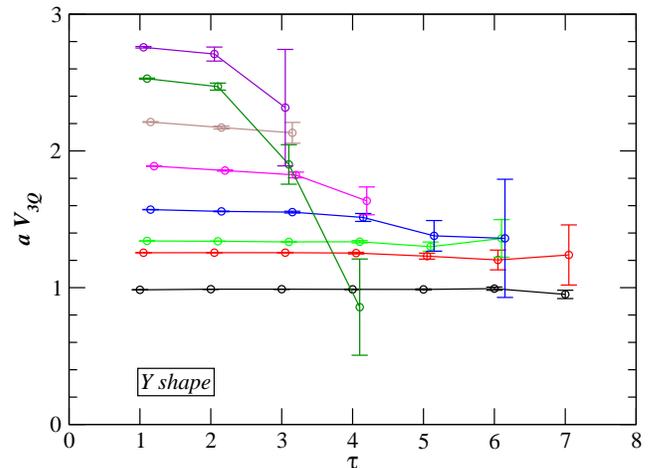}
\centering\includegraphics[width=8.4cm,clip=true]{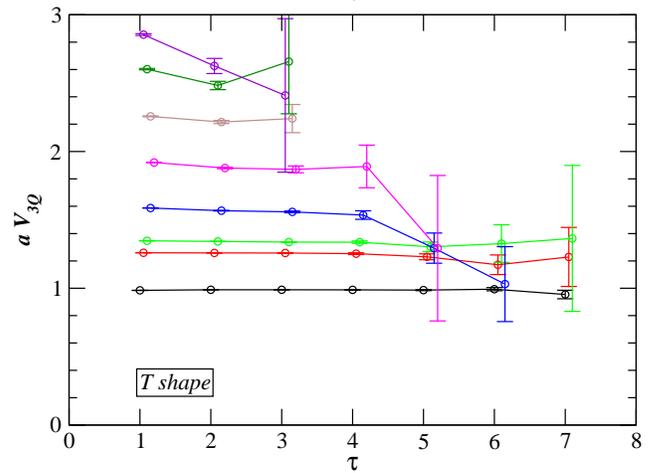}
\centering\includegraphics[width=8.4cm,clip=true]{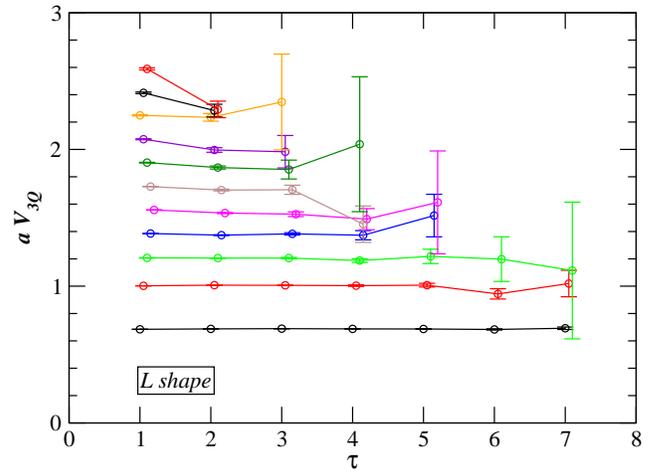}
\caption{Effective static quark potential for 10-sweep smeared Y-, T-
  and L-shape sources.  Some points are offset horizontally to
  increase readability.  From bottom up, the lines correspond to
  shapes 1 through 8 of Table \ref{tab:coord}.}
\label{plat}
\end{figure}

Our results for the various quark positions and source shapes are
shown in Fig. \ref{plat}.  All small shapes are stable against noise
over a long period of time evolution and even some of the largest
shapes show some stability before being lost into the noise.

Robust plateaus are revealed for the first four quark positions of
Table~\ref{tab:coord} for the T and Y shape sources.  This suggests
the ground state has been isolated and indeed the four lowest
effective potentials of the T- and Y-shape sources agree.  This result
was foreseen in the qualitative analysis where Figs.~\ref{Ttube3} and
\ref{Ytube3} for the T- and Y-shape sources respectively displayed the
same correlations between the action density and the quark positions.

Conversely, the disagreement between Figs.~\ref{Ttube7} and
\ref{Ytube7} indicates the ground state has not been isolated in one
or possibly both cases.  Indeed the nontrivial slopes of the seventh
effective potentials of Fig.~\ref{plat} for the Y- and T-shape sources
confirm this.  On the other hand, the curves are sufficiently flat to
estimate an effective potential at small values of $\tau$, and given
knowledge of the node position from our qualitative analysis, one can
make contact with models for the effective potential.

The expected $\vec r$ dependence of the baryonic potential is
\cite{Alexandrou:2002sn,Takahashi:2002bw}
\begin{equation}
V_{3Q} = \frac{3}{2}V_0 -\frac{1}{2}\sum_{j<k} \frac{g^2 C_F}{4\pi r_{jk}}
+\sigma L \, ,
\label{potform}
\end{equation}
where $C_F =4/3$, $\sigma$ is the string tension of the $q\bar{q}$
potential and $L$ is a length linking the quarks.  There are two
models which predominate the discussion of $L$; namely the $\Delta$
and Y ans{\"a}tze.

In the $\Delta$-ansatz, the potential is expressed by a sum of two
body potentials \cite{Alexandrou:2002sn}. In this case $L=L_\Delta /2
=3\langle d_{qq} \rangle /2$ where $L_\Delta$ is the sum of the
inter-quark distances. In the Y-ansatz
\cite{Takahashi:2002bw,Ichie:2002mi}, $L=L_Y = 3\langle r_s\rangle$ is
the sum of the distances of the quarks to the Fermat point.

\FloatBarrier

\begin{figure}
\centering\includegraphics[width=8.5cm,clip=true]{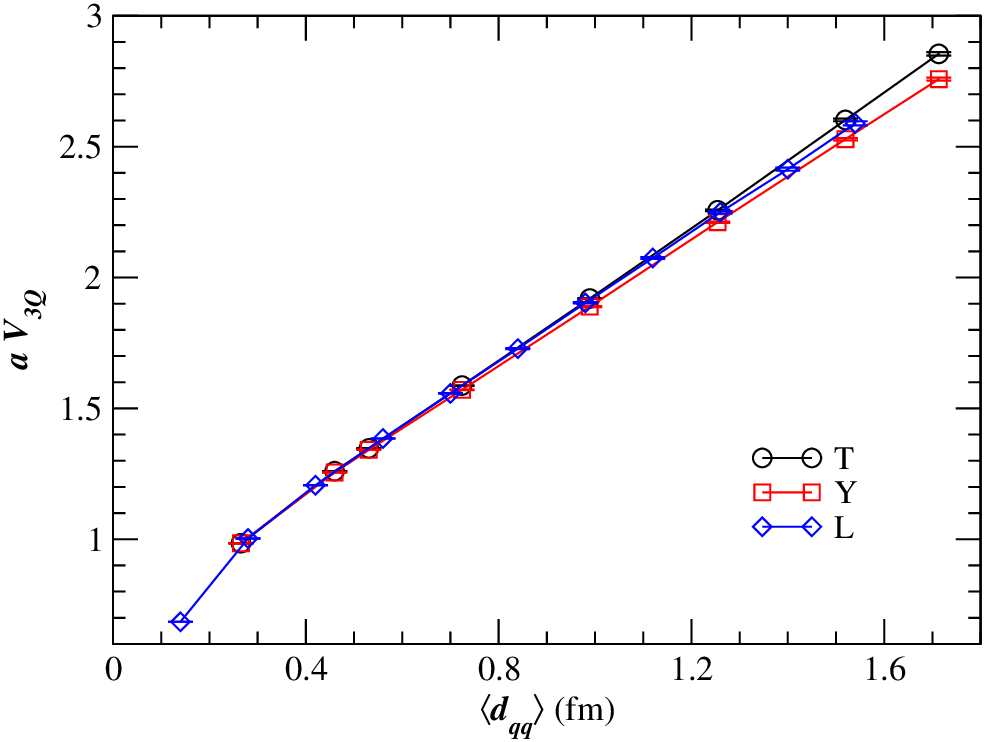}
\caption{Effective potential at $\tau=1$ for various 10-sweep-smeared
  source shapes plotted as a function of the inter-quark separation
  $\langle d_{qq}\rangle= L_\Delta/3$.}
\label{pot-dqq}
\centering\includegraphics[width=8.5cm,clip=true]{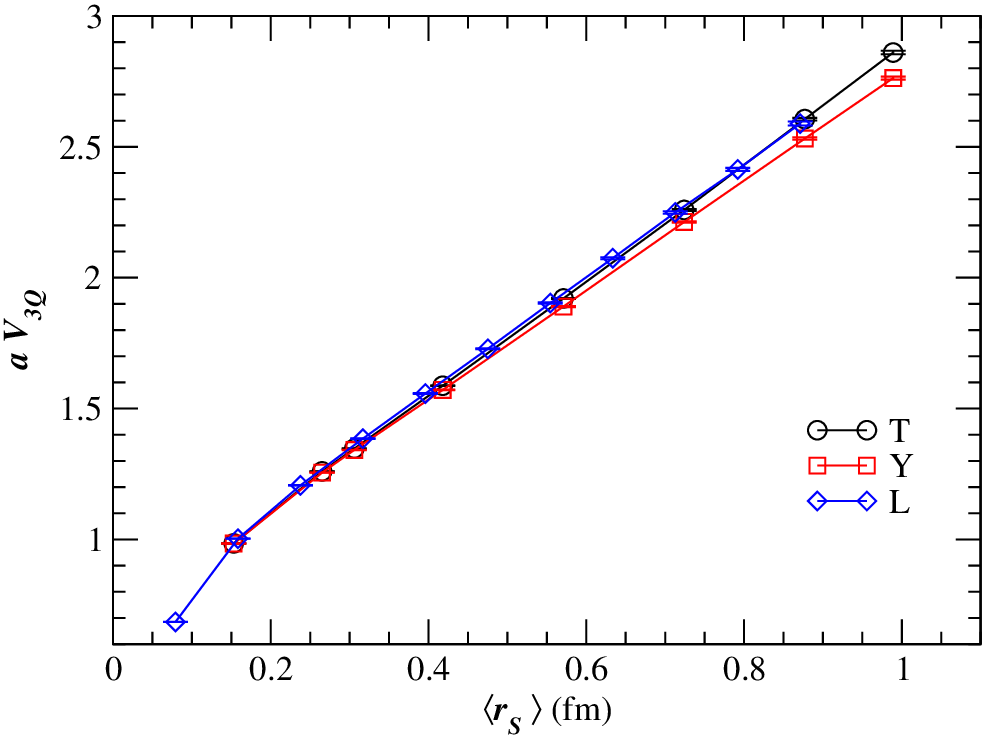}
\caption{Effective potential at $\tau=1$ for various 10-sweep-smeared
  source shapes plotted as a function of the average distance to the
  Fermat point $\langle r_s\rangle= L_Y/3$.}
\label{pot-rs}
\centering\includegraphics[width=8.5cm,clip=true]{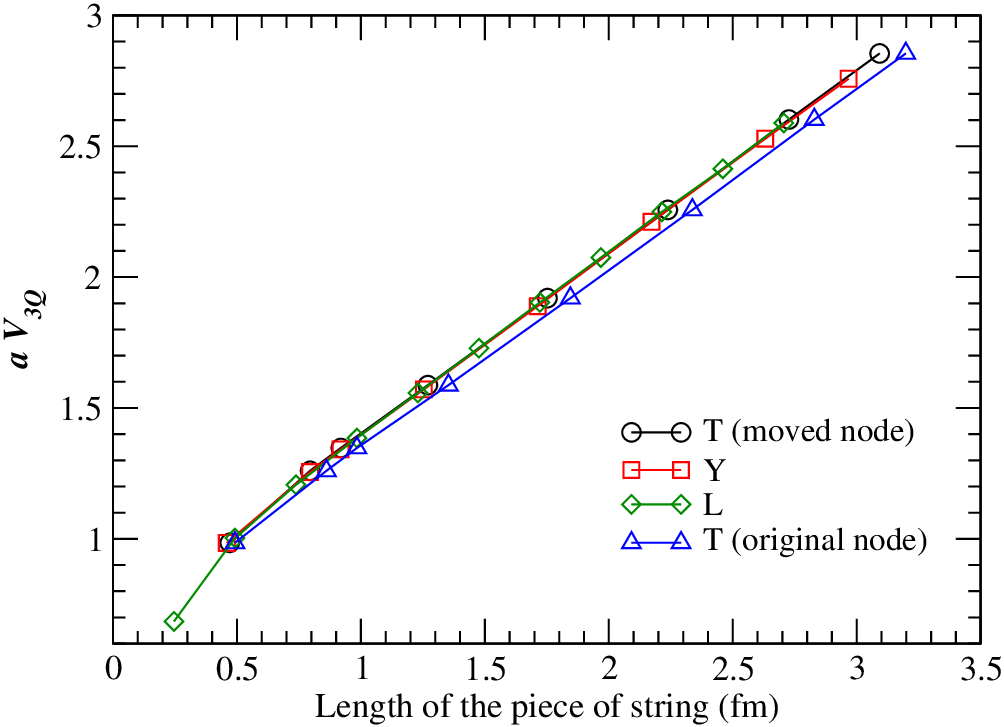}
\caption{Effective potential at $\tau=1$ for various 10-sweep-smeared
  source shapes plotted as a function of the length of string
  connecting the quarks to the node location.}
\label{pot-ls}
\end{figure}

In Figs.~\ref{pot-dqq} and \ref{pot-rs} we show the potentials from
the different shapes at $\tau=1$ as functions of the inter-quark
separation $\langle d_{qq} \rangle$ and the average distance to the
Fermat point $\langle r_s \rangle$ respectively.  In the case of the
L-shape source, it is impossible to distinguish between the two ansatz
from these figures as there is a linear relation between $\langle
d_{qq} \rangle$ and $\langle r_s \rangle$.  However, this is not the
case for the T- and Y- shape sources.  A linear or almost linear
behavior for the long-distance potential is observed for
Fig. \ref{pot-dqq} considering $\langle d_{qq} \rangle$ and similarly
in Fig.~\ref{pot-rs} using $\langle r_s \rangle$.  However, neither of
these methods for expressing the inter-quark separation are capable of
collapsing the potentials arising from T- Y- or L- shape sources to a
single curve.

In Eq. \ref{potform}, $L$ is meant to be the value of a length of
string connecting the quarks (times a geometric constant for the
$\Delta$ ansatz). 
% 
% While Fig. \ref{pot-rs} suggests our potential
% follows a law that is closer to the Y-ansatz, it does not 
% %closely
% follow it exactly.
% However, as 
Figs.~\ref{Tshape7} to \ref{Lshape10} indicate that the actual node of
the flux tube formation is not necessarily in the initial location
shown in Figs.~\ref{tpath} to \ref{ypath} nor, in the case of the T
and L shapes, at the Fermat point.  Using our knowledge of the node
position in the observed state (which may not have relaxed to the
ground state), we can determine the length of string required to
connect the quark locations to the observed node position.

In the case of the Y-shape source, both the observed node and the
source node are very close to the Fermat point.  The difference in the
total distance from the quark to the node is at most a few percent
compared to $L_Y$.
%However, 
In the case of the T-shape source the observed node has moved
by one lattice unit from the top of the T towards the Fermat point.
In the smallest shape this means the node is extremely close to the
Fermat point and that it is in the same location as the Y-shape node.
In the case of the largest T-shape source considered in
Table~\ref{tab:coord}, the node may have moved more than one lattice
unit.

In Fig.~\ref{pot-ls} we plot the potential of the various source
shapes and sizes as a function of the total length of string connecting the
quarks to the observed node.  The length of string is computed as
follows:
\begin{itemize}
\item{For the T shape, it is computed as the total distance from the
  observed node moved one lattice unit from the top of the T towards
  the Fermat point.}
\item{For the Y shape, it is taken to be equal to $L_Y$.}
\item{For the L shape, it is taken to be $2\ell$ for the corresponding
shape.}
\end{itemize}
Using this measure on the $x$-axis, all the data points fall close to
a single curve as displayed in Fig.~\ref{pot-ls}.  The three curves
for T-, Y- and L-shape sources with the node movement included are
extremely close to each other.  The outstanding curve labeled ``T
(original node)'' emphasizes the importance of including node movement
identified by the three-point correlation function.  This observation
favours the Y-ansatz as providing the correct understanding of the
nontrivial, nonperturbative dynamics of QCD.

The asymptotic part of all the curves follows a straight line as
suggested by Eq. \ref{potform} and we can extract a string tension
from this part.  From the slope of the asymptotic part of these curves
the string tension is $\sigma \approx 1.11\pm 0.01\ \text{GeV}/\text{fm}$.
This is somewhat different from the results
from Takahashi \textit{et al.}  \cite{Takahashi:2002bw} of $0.89\
\text{GeV}/\text{fm}$ and also from the value we used to fix the
lattice spacing ($0.981\ \text{GeV}/\text{fm}$).  On the other hand,
Ref.~\cite{deForcrand:2005vv} established that the Y-law describes the
asymptotic potential with the same string tension  as the
$q$-$\overline q$ potential.

Since the Y shape source is already in a configuration providing
access to the shortest string length,  we must conclude that other
aspects of the flux tube distribution have yet to evolve to become the ground
state.  Aspects including the flux-tube radius and the extent to
which vacuum field fluctuations are suppressed inside the flux tube
are examined in the following sections.

\section{SIMULATION RESULTS WITH 30 SWEEPS SMEARING}
\label{results30}

Given that simple Euclidean time evolution to the ground state is
difficult for large quark separations, we turn our attention to
improving the source by adjusting the amount of spatial-link smearing
used in the source construction.

For example, Takahashi \textit{et al.} \cite{Takahashi:2002bw} used a
large amount of spatial smearing in their studies compared to our
first study of 10 smearing-projection sweeps.  We have found
that an increase in the amount of smearing leads to a reduction in the
value of the potential for large quark separations while the
smaller separations ($r_s<0.5\ \text{fm}$) remain unchanged. This
demonstrates that we have indeed isolated the ground state for the
smaller shapes after 10 smearing sweeps, but not for the larger
shapes.

\begin{figure}
\centering\includegraphics[width=8.5cm,clip=true]{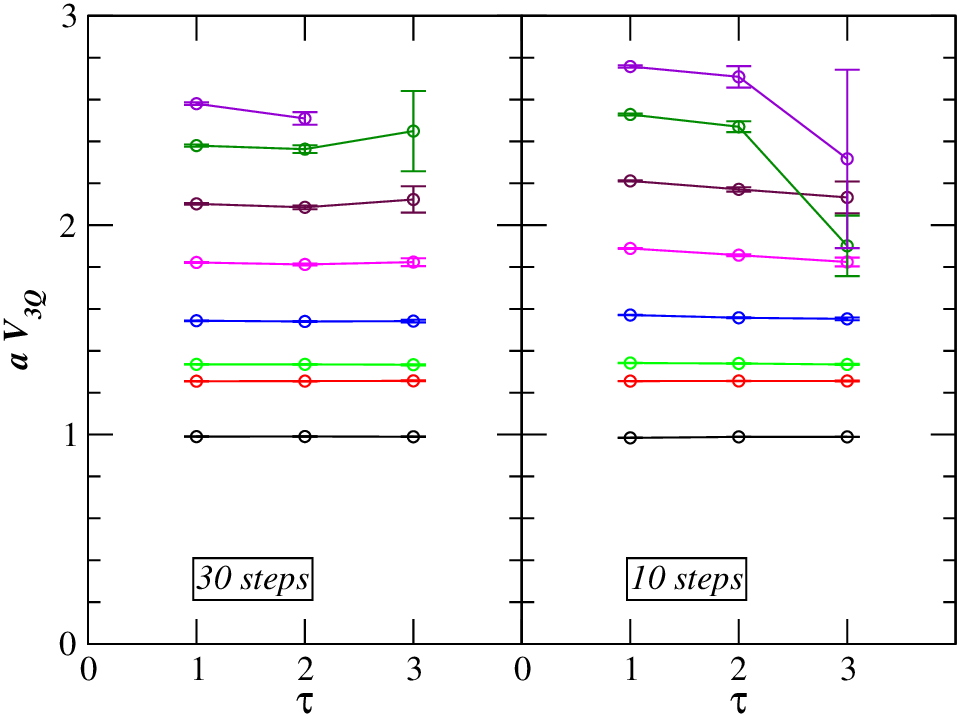}
\caption{Comparison of the potential of the Y shapes for 10 and 30 steps
smearing.  From bottom up, the lines correspond to
  shapes 1 through 8 of Table \ref{tab:coord}.}
\label{ysmear-comp}
\centering\includegraphics[width=8.5cm,clip=true]{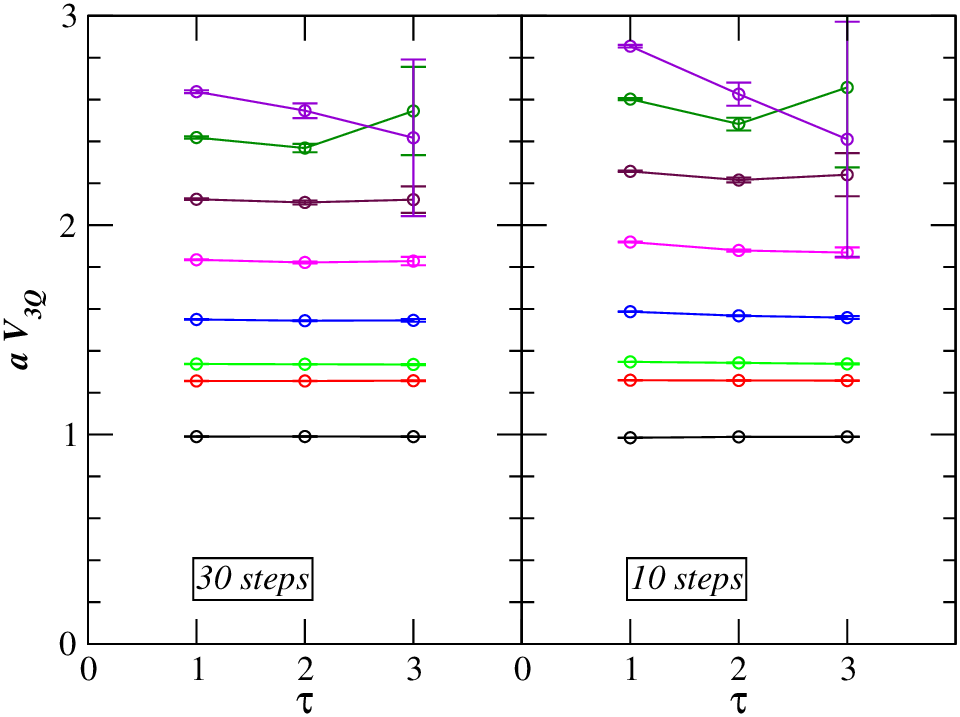}
\caption{Comparison of the potential of the T shapes for 10 and 30 steps
smearing.  From bottom up, the lines correspond to
  shapes 1 through 8 of Table \ref{tab:coord}.}
\label{tysmear-comp}
\end{figure}

We therefore investigated systematically the effect of the smearing on
the string tension.  We find that the string tension decreases for
increased smearing until it reaches a plateau at $30$ steps of
smearing.  The resulting value for $\sigma = 0.97\pm 0.01\
\text{GeV}/\text{fm}$ is in agreement with the value used
to fix the lattice spacing, $\sigma = 0.98$ GeV/fm.

The increase in smearing provides excellent plateaus in the effective
potential of the Y-shape source at the earliest times.  This can be
seen in Fig.~\ref{ysmear-comp} which provides a side-by-side
comparison of 30 and 10 smearing sweeps for the effective potential
obtained from the first three values of the time parameter $\tau$.
Improved plateau behavior and lower effective potentials are observed
for all separations.

\begin{figure}
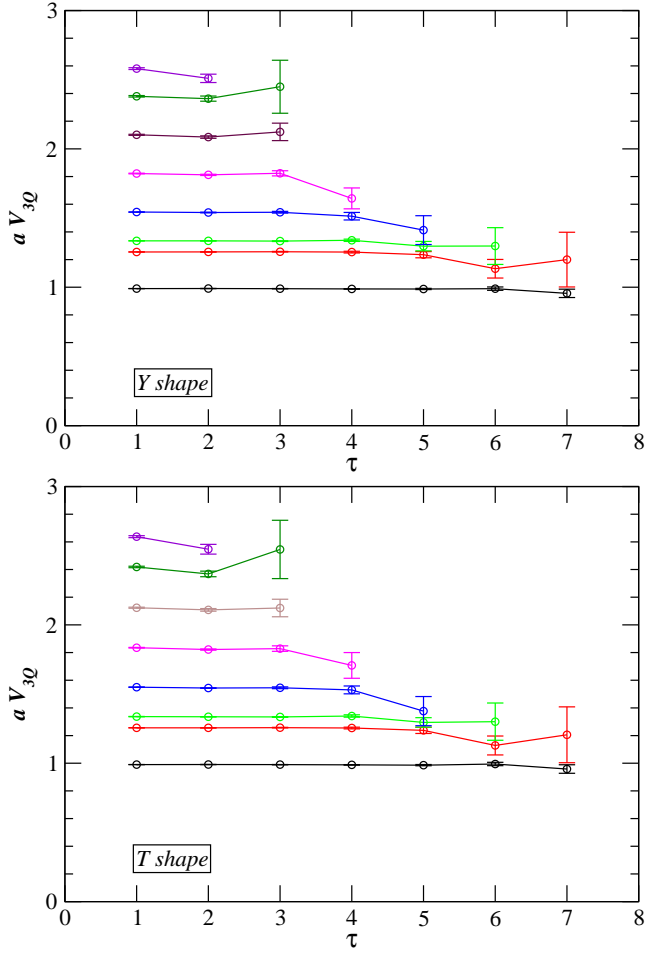

\centering\includegraphics[width=8.5cm,clip=true]{article-Y-plateau-cd8-s30}
\centering\includegraphics[width=8.5cm,clip=true]{article-T-plateau-cd8-s30}
\caption{Study of the stability of the effective potential over
  Euclidean time for 30-sweep smeared Y and T shapes.  From bottom up,
  the lines correspond to shapes 1 through 8 of Table
  \ref{tab:coord}.}
\label{plat-s30}
\end{figure}

Similar results are observed for the T-shape source as illustrated in
Fig.~\ref{tysmear-comp}.  Figure~\ref{plat-s30} shows the complete
time evolution of the potential of Y and T shapes for the 30 sweep
smearing results.  While it is clear that the ground state has not yet
been reached for the largest T-shape source separations, it will be
fascinating to observe how the T-shape source flux-tube distribution
has evolved to produce a much lower effective potential.

\begin{figure}
\centering\includegraphics[height=6.2cm,clip=true]{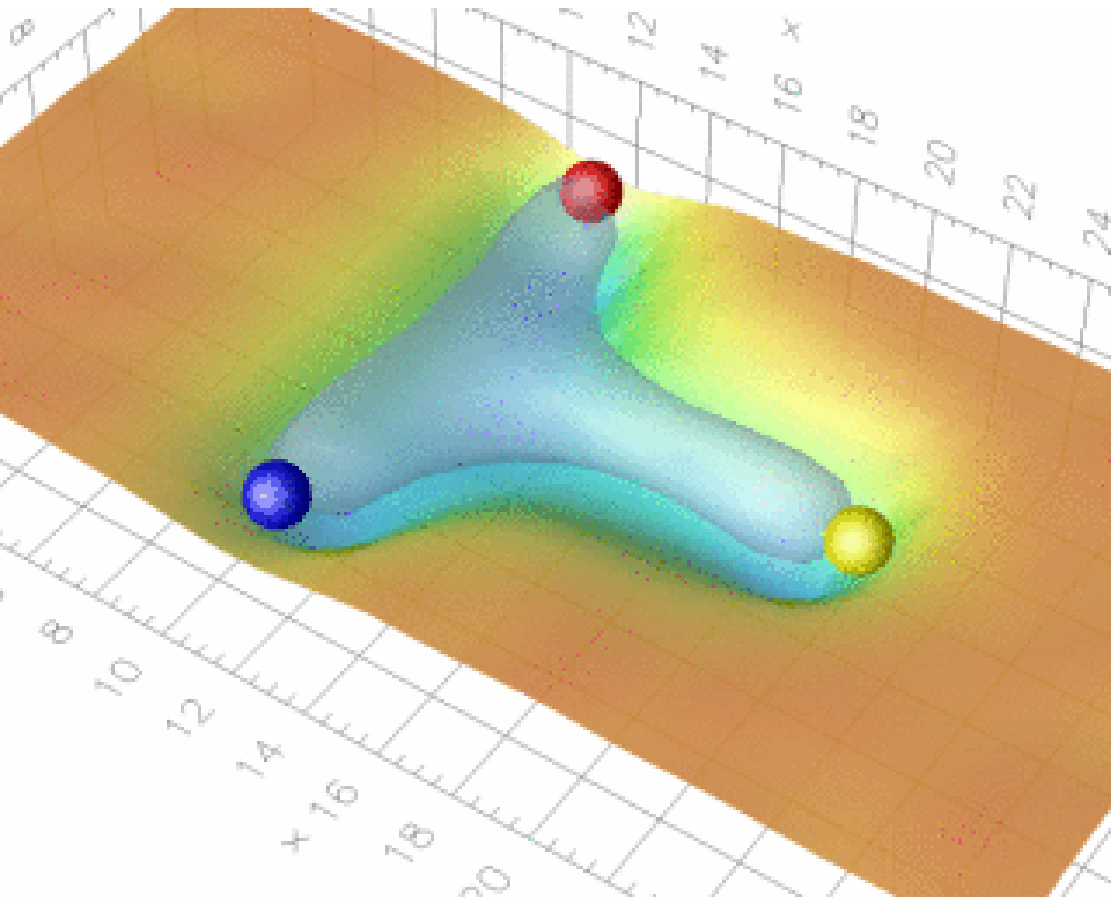}
\caption{Expulsion of gluon-field fluctuations from the region of
  static quark sources illustrated by the spheres.  An isosurface of
  $C(\vec{y})$ is illustrated by the translucent surface.  A
  surface plot (or rubber sheet) describes the values of $C(\vec{y})$
  for $\vec y$ in the quark plane, $(y_1, y_2, 0)$.  Results are for a
  30-sweep smeared T-shape source with quark positions as in the seventh
  configuration of Table \ref{tab:coord}. The maximum expulsion is 6.4\%
  and the isosurface is set to 4.4\%.}
\label{Ttube7-s30}
\centering\includegraphics[height=6.2cm,clip=true]{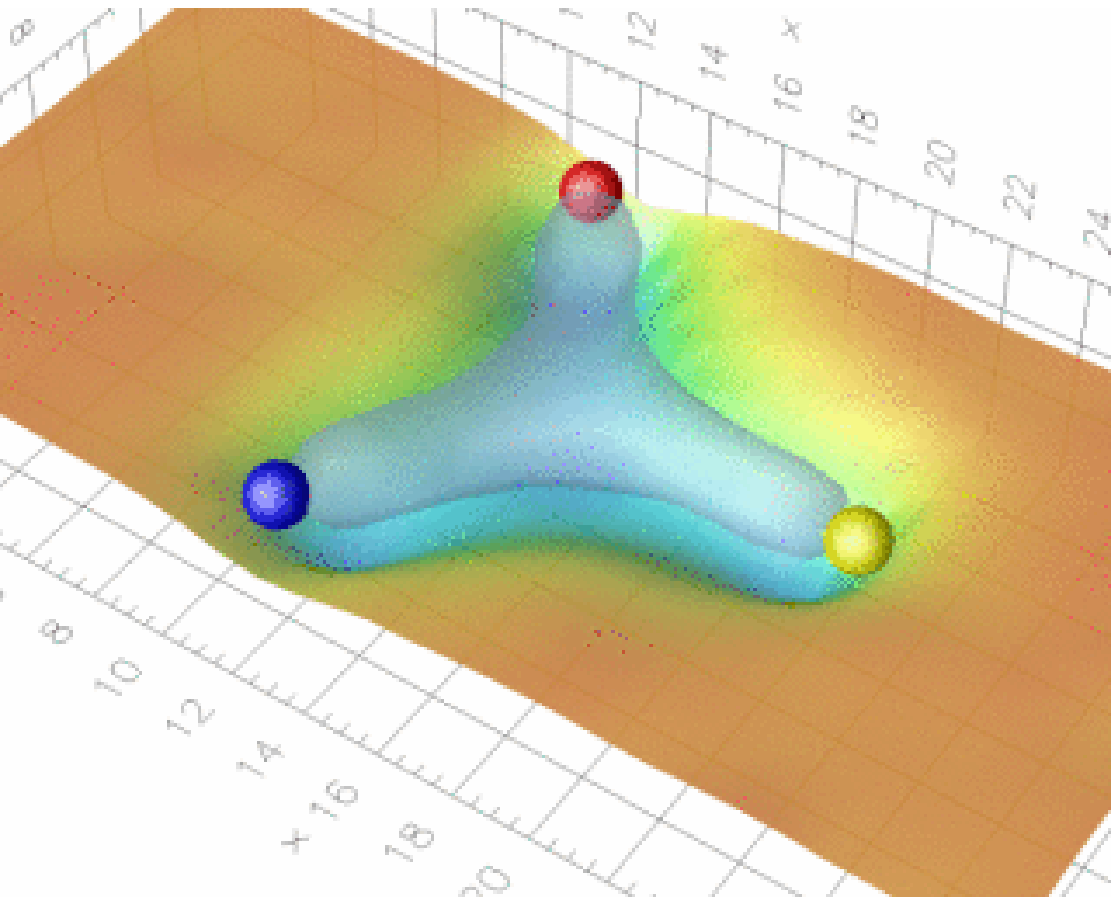}
\caption{Isosurface and surface plot of $C(\vec{y})$ for a 30-sweep
   smeared Y-shape source with quark positions as in the seventh
   configuration of Table \ref{tab:coord}. The maximum expulsion
   is 6.4\% and the isosurface is set to 4.4\%. Further details are described
   in the caption of Fig.~\protect\ref{Ttube7-s30}.}
\label{Ytube7-s30}
\end{figure}

The gluon-field distributions for the smaller source shapes are
largely unchanged by the shift from 10 to 30 steps of smearing and
we do not reproduce them here.  However interesting results are
revealed in the larger source shapes.  The seventh configuration of
Table~\ref{tab:coord} is depicted in Figs.~\ref{Ttube7-s30} and
\ref{Ytube7-s30} for the T- and Y-shape sources respectively.

\begin{figure}
\centering\includegraphics[width=8.5cm,bb= 0 52 320 240,clip=true]{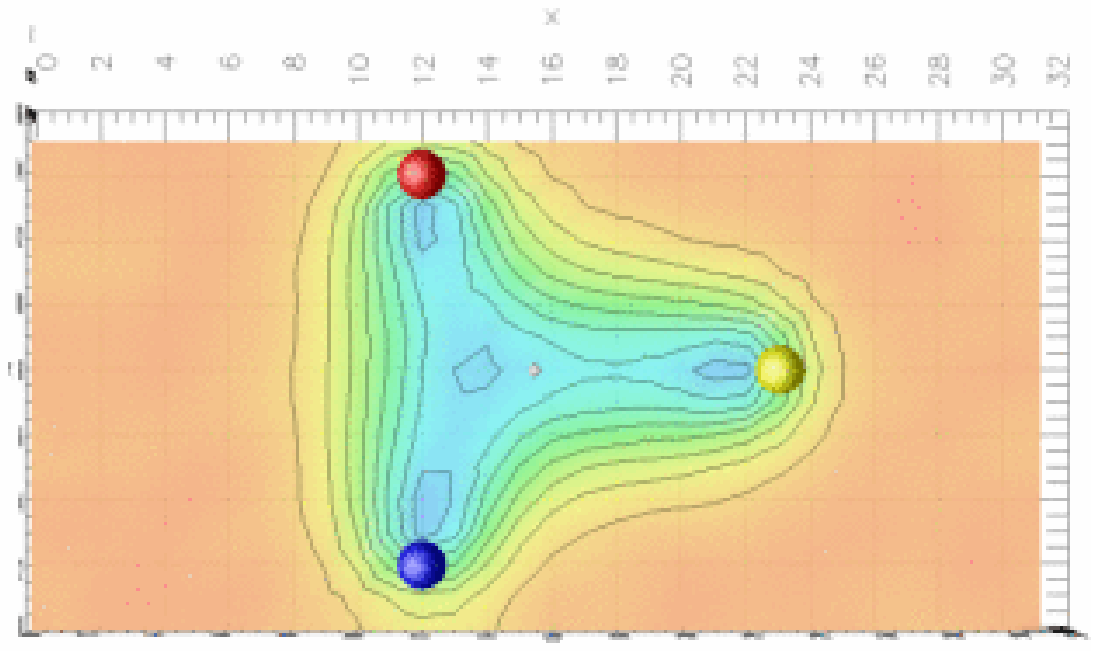}
\caption{Contour plot of $C(\vec{y})$ for $\vec y$ in the quark plane,
  $(y_1, y_2, 0)$, for a 30-sweep smeared T-shape source with quark
  positions as described in the seventh configuration of Table
  \ref{tab:coord}.  The white diamond shows the position of the Fermat
  point.}
\label{Tshape7-s30}
\centering\includegraphics[width=8.5cm,bb= 0 52 320 240,clip=true]{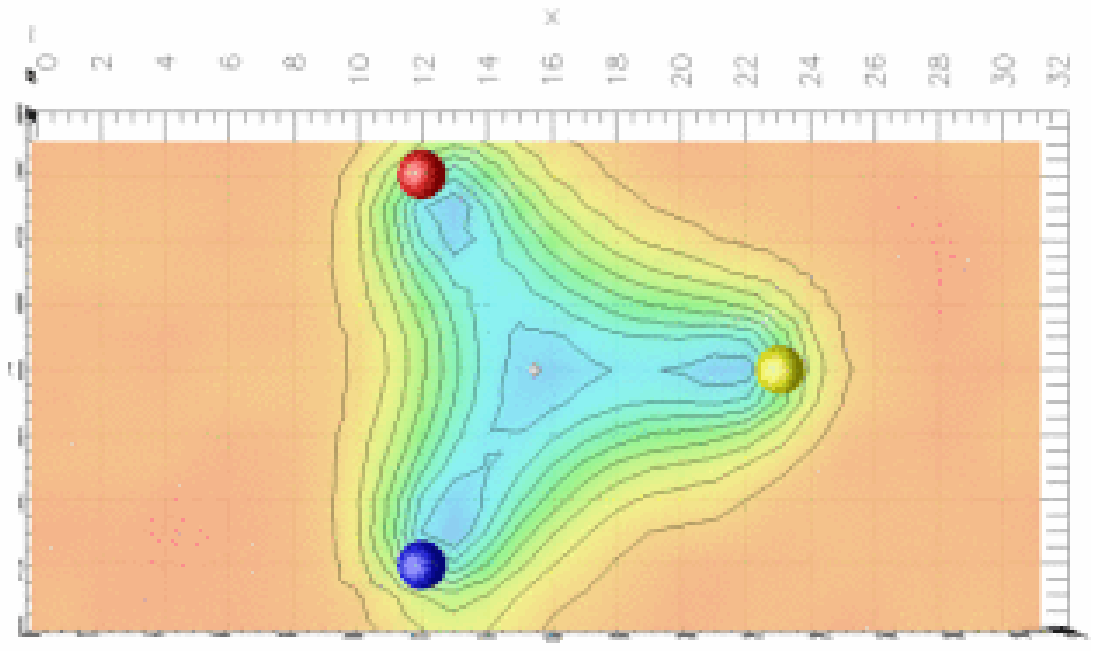}
\caption{Contour plot of $C(\vec{y})$ for $\vec y$ in the quark plane,
  $(y_1, y_2, 0)$, for a 30-sweep smeared Y-shape source with quark
  positions as described in the seventh configuration of Table
  \ref{tab:coord}.  The white diamond shows the position of the Fermat
  point.}
\label{Yshape7-s30}
\end{figure}

The observed flux distribution of the T-shape source in
Fig.~\ref{Ttube7-s30} displays significant evolution towards the
observed Y-shape of Fig.~\ref{Ytube7-s30} (and not towards a $\Delta$
shape).  The node connecting the quarks in the T-shape 
source has now moved two lattice units in the direction of the Fermat 
point rather than only a single unit as in the 10-sweep smearing case.  
While this is not so obvious in Fig.~\ref{Ttube7-s30} it can be clearly 
observed in the contour plot of Fig.~\ref{Tshape7-s30} and directly 
compared to Fig.~\ref{Tshape7}.  It is approaching the observed Y-shape 
source result of Fig.~\ref{Yshape7-s30}.  In contrast, the Y-shape source
flux-tube distribution is qualitatively unchanged, as can be seen in
Figs. \ref{Ytube7-s30} and \ref{Yshape7-s30} compared to
Figs. \ref{Ytube7} and \ref{Yshape7}.

The isosurface of Fig.~\ref{Ytube7-s30} and contour plot of
Fig.~\ref{Yshape7-s30} provide our best determination of the
distribution of gluon flux tubes within the ground state of the
static-quark baryon.

\begin{figure}
\centering\includegraphics[width=8.5cm,clip=true]{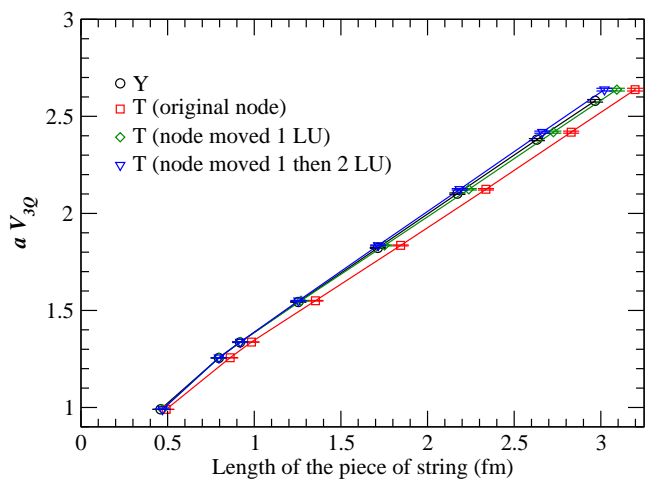}
\caption{Effective potential at $\tau=1$ for various 30-sweep-smeared
  source shapes plotted as a function of the length of string
  connecting the quarks to the node location.}
\label{pot-ls-s30}
\end{figure}

Motivated by our earlier discovery and knowledge of the precise
location of the flux-tube node, we plot the effective potential as a
function of the total length of the piece of string connecting the
quarks to the node in Fig.~\ref{pot-ls-s30}.  The length of the piece
of string is now computed as follows:
\begin{itemize}
\item{For the Y shape, it is taken to be equal to $L_Y$.}
\item{For the T shape, we first compute the total distance from a node moved
one lattice unit from the base of the T.}
\item{We also consider a curve for the T shape where the node is moved
one lattice unit from the top of the T for small shapes (1 to 5 in 
Table \ref{tab:coord}) and two lattice units for the larger shapes.}
\end{itemize}
These results which converge very well to a single curve are
contrasted by a curve for the T shape where the length of string is
computed from the source-node position located at the top of the T
(original node).

As can be seen in Fig. \ref{pot-ls-s30} the latter effective potential
%is far 
differs from the effective potential of the Y shapes.  On the other
hand, the two curves for the T-shape source with the node moved one or
two lattice steps are very close together with the 
curve for the Y shapes.  From
this figure alone it is quite hard to decide which of the two moved T
shapes is the closest to the Y shape.  The following section providing
a quantitative analysis of various cross sections of $C(\vec y)$ will
resolve this issue.

\section{FLUX-TUBE PROPERTIES}
\label{tubeProps}

Here the flux-tube and node properties as revealed by the three-point
correlation function analysis in $C(\vec y; r_1, r_2, r_3; \tau)$ are
quantitatively examined by considering several cross sections of $C$.
In all cases $\vec y$ is constrained to the plane of the quarks $(y_1,
y_2, 0)$ and it is convenient to refer to these components as $(x, y,
0)$ where $x$ is the coordinate of the long direction.  For the
remainder of this section the origin of the coordinate system is placed
at the original position of the source node connecting the quarks in the
computations.

\begin{figure}
\centering\includegraphics[clip=true]{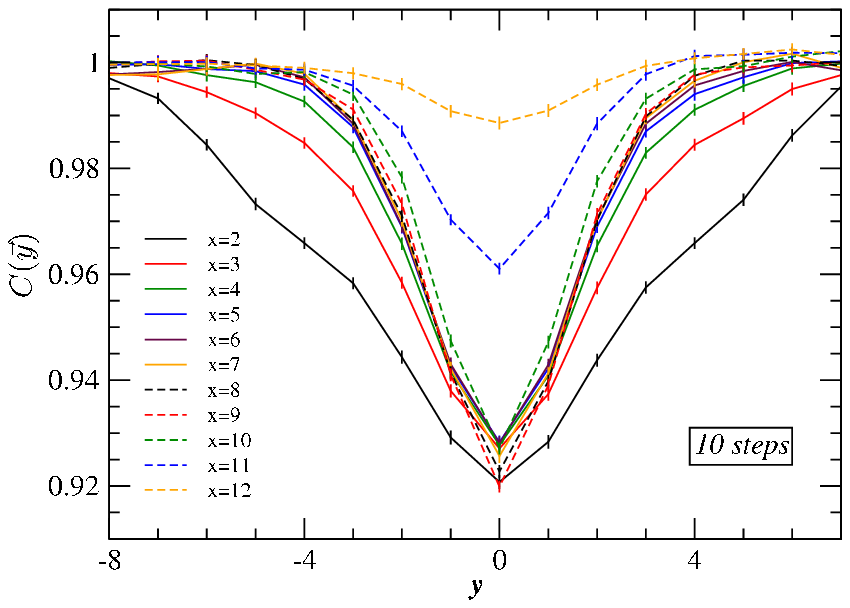}
\centering\includegraphics[clip=true]{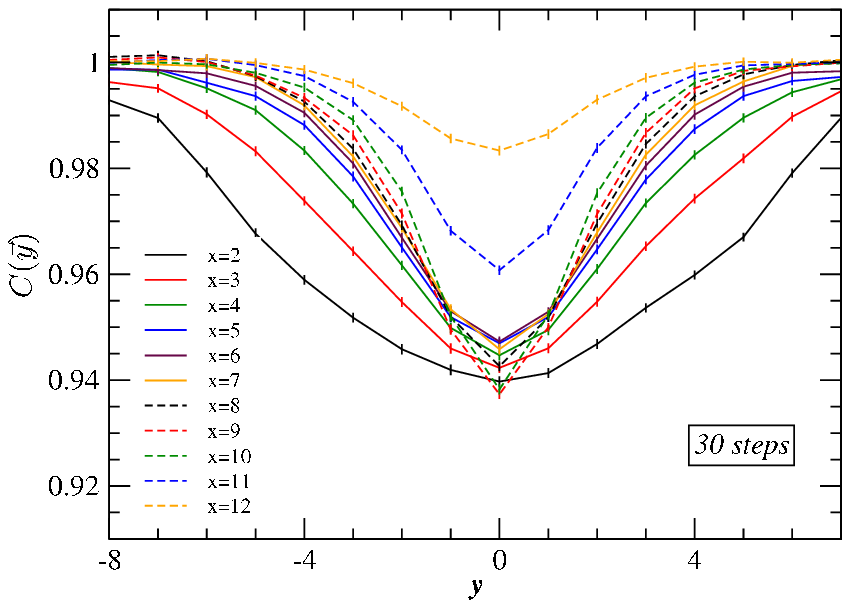}
\caption{Value of $C(x,y,0)$ for the seventh T-shape source of
  Table~\ref{tab:coord} as a function of $y$ (the short direction of
  the quark plane) for various fixed values of $x$ (the long
  direction).  Results from sources constructed from 10 (top) and 30
  (bottom) sweeps of APE spatially-smeared links are compared.  In
  these plots the origin is placed at the position of the node in the
  source.}
\label{flux-t7}
\end{figure}

We begin with a study of $C(x, y, 0; \tau)$ with $\tau = 2$ for a fixed
value of $x$.  Larger values of $\tau$ are examined below.
In Fig.~\ref{flux-t7} we show the cross-section of the observed flux
tube for the seventh T-shape source.  Each curve corresponds to a
fixed value of $x$.  At $x=2$ we are at the edge of the observed node
position.  At $x=12$ we are past the quark in the long direction which
is located at $x=11$.  Results for both 10- and 30-sweep smeared
T-shape sources are displayed.

\begin{figure}
\centering\includegraphics[clip=true]{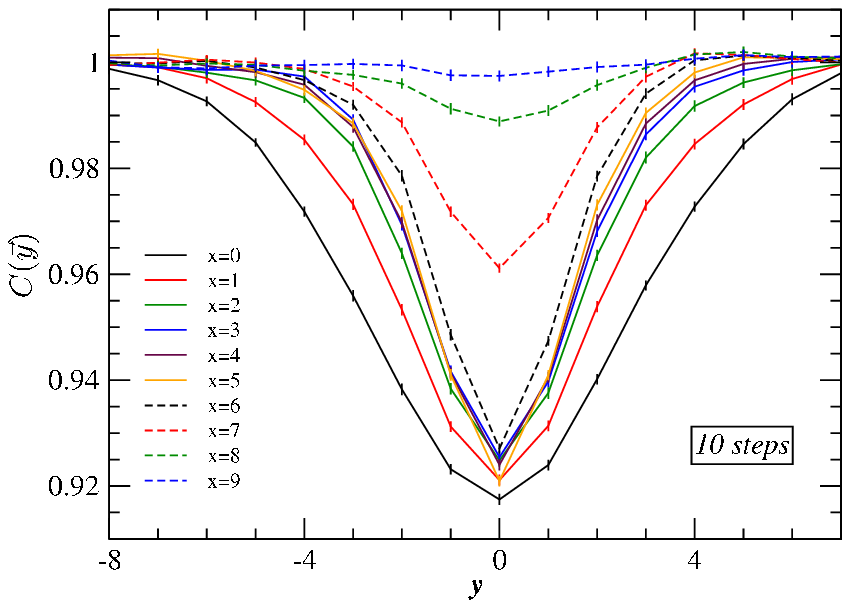}
\centering\includegraphics[clip=true]{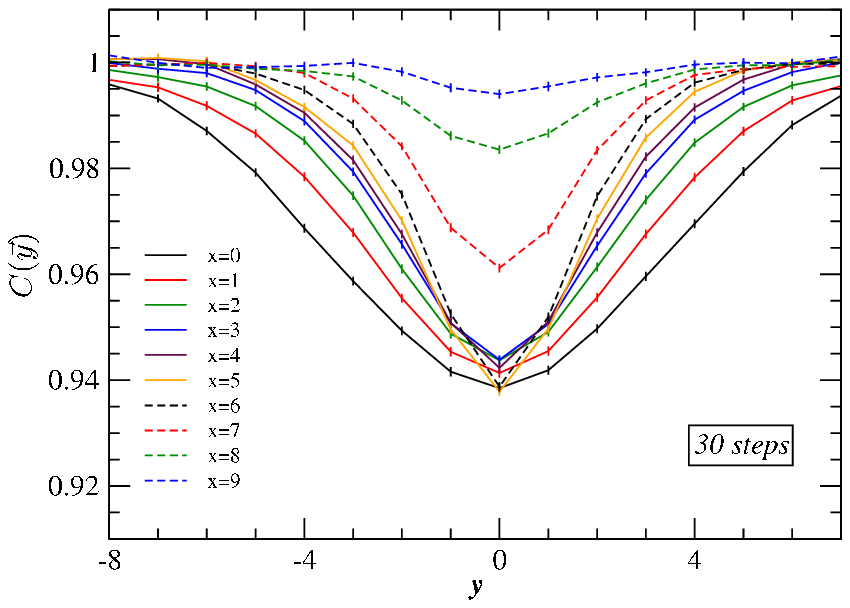}
\caption{Value of $C(x,y,0)$ for the seventh Y-shape source of
  Table~\ref{tab:coord} as a function of $y$ (the short direction of
  the quark plane) for various fixed values of $x$ (the long
  direction).  Results from sources constructed from 10 (top) and 30
  (bottom) sweeps of APE spatially-smeared links are compared.  In
  these plots the origin is placed at the position of the node in the
  source.}
\label{flux-y7}
\end{figure}

Similarly, in Fig.~\ref{flux-y7} we show the cross-section of a flux
tube observed for the seventh Y-shape source.  Again, each curve
corresponds to a fixed value of $x$.  At $x=0$ we are close to the
position of the node.  At $x=9$ we are past the quark in the long
direction which is located at $x=7$ in this case.

While the 10-sweep-smearing computations display stronger vacuum-field
action suppression, the 30-sweep-smearing results display a larger
flux-tube diameter.  The latter also display a Gaussian shape in the
vacuum-field suppression.  We can also see enhanced action suppression
near the location of the node and just before the position of the
quark.  While the latter feature is suppressed to some extent by our
consideration of a highly-improved action density, the greater action
suppression at the node is physically interesting.  The convergence of
the curves in the intermediate regime reflects the formation of a well
defined flux tube.

For the T shape, the tube is reasonably well defined for $x=4$ through
$7$ in both the 10- and 30-sweep-smeared sources.  For the Y shape
similar features appear for $x=2$ to $5$, again for both source
smearings.  We emphasize that it is the 30-sweep Y-shape source that
provides the best overlap with the ground-state potential, as it is
this source that provides the smallest and flattest effective
potential.  Therefore, the 30-sweep Y-shape source provides the best
determination of the actual shape of the flux tube within the
ground-state baryon system.

\begin{figure}
\centering\includegraphics[clip=true]{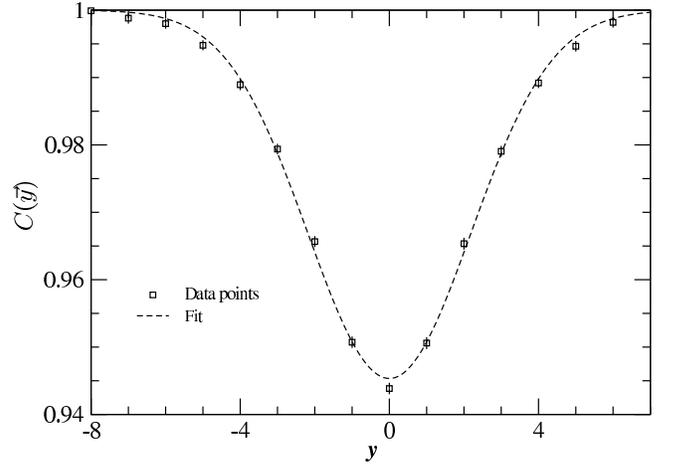}
\caption{Example fit of a Gaussian to the flux-tube cross-section as
per Table \ref{flux-fit}.  The data points are for the seventh Y shape
with 30 steps smearing as in the lower part of Fig.~\ref{flux-y7} but
only for $x=3$.}
\label{fit-flux-ys30}
\end{figure}

We can estimate the radius of the flux-tube by fitting a function of
the form $1-A\, e^{-y^2/r^2}$ to selected curves in
Figs.~\ref{flux-t7} and \ref{flux-y7}.  For the T shape we choose to
fit $x=5$ and similarly we select $x=3$ for the Y shape.  All the
non-linear fits hereafter have been done using the Levenberg-Marquardt
method \cite{NRC92}.  The results are summarised in
Table~\ref{flux-fit}. The fits are very good as illustrated by
Fig.~\ref{fit-flux-ys30}.

\begin{table}
\caption{Values of the fit parameters of the function $1-A\,
  \exp(-y^2/r^2)$ to the curves $x=5$ of Fig.~\ref{flux-t7} and $x=3$
  of Fig.~\ref{flux-y7}.  $r$ is reported in lattice units (LU) and
  fm.  The last column indicates the area under $C(\vec{y})=1$ for the
  fitted curve in units of relative-depth times LU.}
\label{flux-fit}
\begin{ruledtabular}
\begin{tabular}{ccccc}
Shape        & $A$        & $r$ (LU)   & $r$ (fm)  & $Ar \sqrt{\pi}$ \\
\noalign{\smallskip}
\hline
\noalign{\smallskip}
T (10 steps) & $0.070(1)$ & $2.3(1)$   & $0.28(1)$ & $0.28(1)$ \\
Y (10 steps) & $0.073(1)$ & $2.2(1)$   & $0.27(1)$ & $0.28(1)$ \\
T (30 steps) & $0.052(1)$ & $3.3(1)$   & $0.41(1)$ & $0.31(1)$ \\
Y (30 steps) & $0.055(1)$ & $3.1(1)$   & $0.38(1)$ & $0.30(1)$ \\
\end{tabular}
\end{ruledtabular}
\end{table}

Our earlier observations about the depth and width of the flux-tubes
for 10 versus 30 source-smearing sweeps are confirmed in
Table~\ref{flux-fit}.  There is a significant correlation between the
smeared-source size and the observed flux-tube radius.  Whereas the
source radius increases by 70\% in going from 10 to 30 sweeps at a
smearing fraction of 0.7, the observed flux-tube radius increases by
only 40\% for the Y-shape source.  This indicates that some Euclidean
time evolution of the flux-tube diameter has occurred in the 10-sweep
smeared case, but not enough to relax to the true ground state
captured by the 30-sweep smeared Y-shape source.  Our best estimate of
the ground-state flux-tube radius from the 30-sweep Y-shape source
is $r = 0.38(3)$ fm, where the uncertainty includes both statistical
errors and the consideration of neighboring flux-tube cross-sections.
Similarly, $A = 0.055(2)$ for the relative vacuum expulsion in the
flux tube of the seventh quark separation.

Finally in the last column we have the cross-sectional area created by
the tube, computed from the fit.  Despite rather different radii and
depths observed for the two different smearing levels, this quantity
is relatively uniform.  The breadth of the 30-sweep-smeared sources
does give rise to a larger expulsion of vacuum field fluctuations in
the flux-tube of the action density distribution, despite the fact
that these sources give rise to a lower static quark potential.

\begin{figure}
\centering\includegraphics[clip=true]{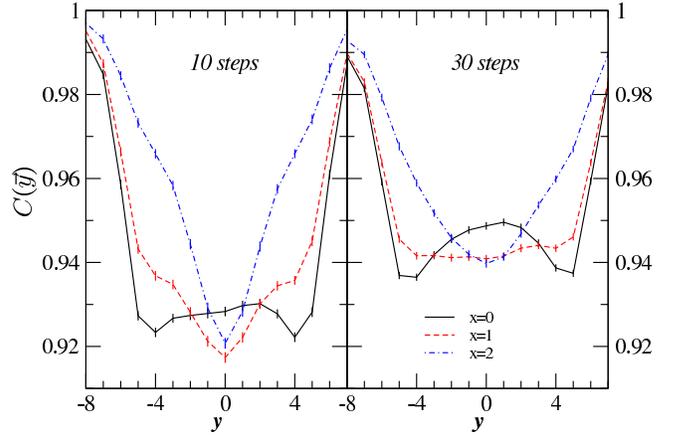}
\caption{Value of $C(x,y,0)$ for the seventh T-shape source of
  Table~\ref{tab:coord} as a function of $y$ (the short direction of
  the quark plane) for various fixed values of $x$ (the long
  direction) near the top of the T-shape source.  Results from sources
  constructed from 10 (left) and 30 (right) sweeps of APE
  spatially-smeared links are compared.  The origin is placed at the
  position of the node in the source.}
\label{t-base}
\end{figure}

The observed flux distribution of the 30-sweep T-shape source in
Fig.~\ref{Ttube7-s30} displays significant evolution towards the
observed Y-shape of Fig.~\ref{Ytube7-s30} (and not towards a $\Delta$
shape).  Further proof of this node displacement is presented in
Fig.~\ref{t-base}.  Here the asymmetry about $x=0$ provides an
estimate of the statistical errors.  In both the 10- and 30-sweep
cases on the $x=0$ line, we can see the response to the quarks on both
sides of the top of the T at $x = \pm 6$ and less suppression of the
vacuum action in the middle.  This indicates the node has moved in
both cases.  However, in the case of 10 sweeps of smearing, the node
is encountered already at $x=1$.  On the other hand with 30 sweeps of
smearing one observes a plateau which shows that the combined effect
of the quarks and the node makes $C(\vec{y})$ constant.  At $x=2$
where we pass through the position of the node, the dip from the node
is clear but it is almost the same depth as the plateau at $x=1$.

\begin{figure}
\centering\includegraphics[clip=true]{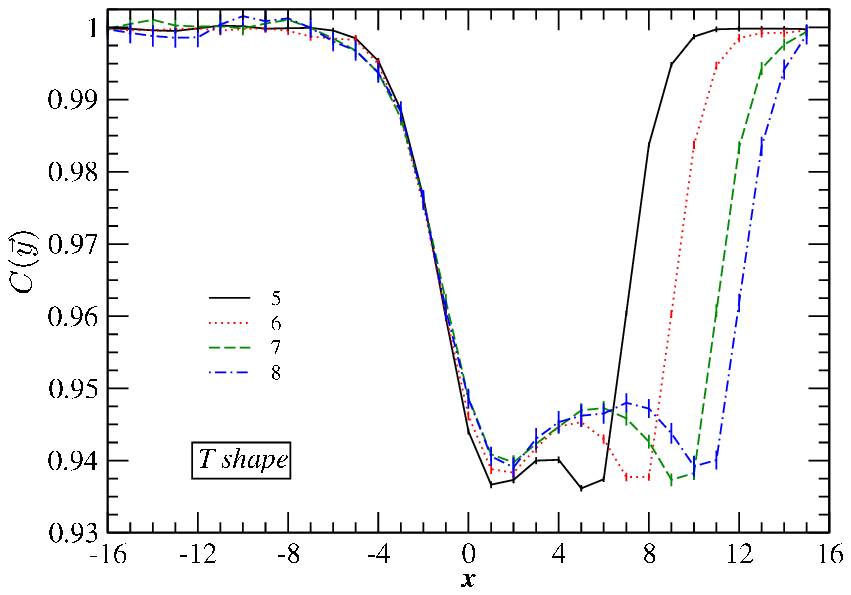}
\centering\includegraphics[clip=true]{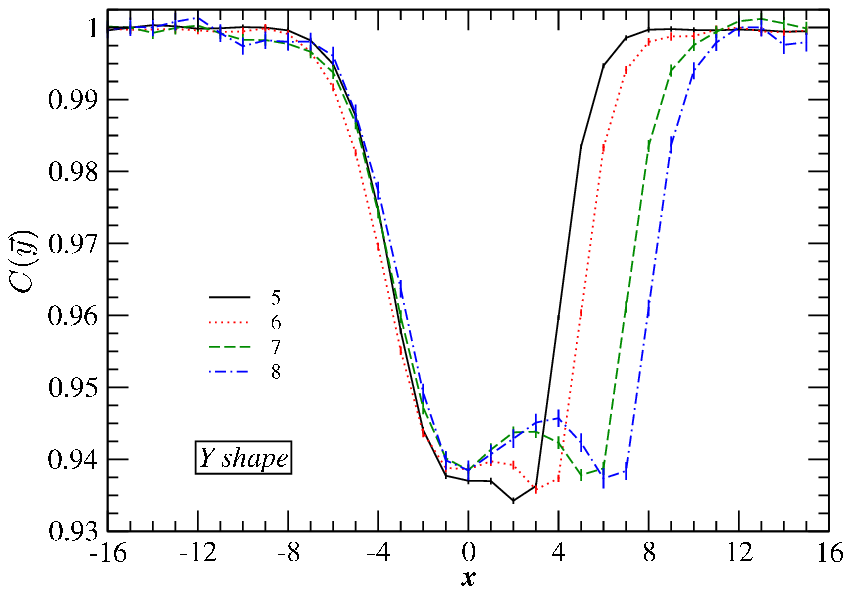}
\caption{Value of $C(x,y,0)$ for a variety of T-shape (top) and
  Y-shape (bottom) sources of Table~\ref{tab:coord} as a function of
  $x$ (the long direction of the quark plane) for $y=0$ such that the
  centre of the node is encountered.  Results are for the 30-sweep
  smeared sources.  In these plots the origin is placed at the
  position of the node in the source.}
\label{nodes}
\end{figure}

Another perspective on the node properties is obtained by plotting the
value of $C(x,y,0)$ along the line where the node and one quark stand
on the long axis, $y=0$.  We show this in Fig.~\ref{nodes} for several
30-sweep T- and Y-shape sources.  In these plots the $x=0$ position
indicates the original position of the node in the source.

In these figures we can typically observe two minima: one for the node
on the left and one inside the quark position on the right. We choose
to plot only the shapes for which these two features are resolved.
For smaller shapes the node and quark peaks are not separable.

For the T shape the node moves first one lattice unit (shape 5) then
two (shapes 6 to 8).  On the other hand, for the Y-shape sources the
node stays at its original location, indicating that the Y-shape
source is already accessing the ground state.

\begin{figure}
\centering\includegraphics[clip=true]{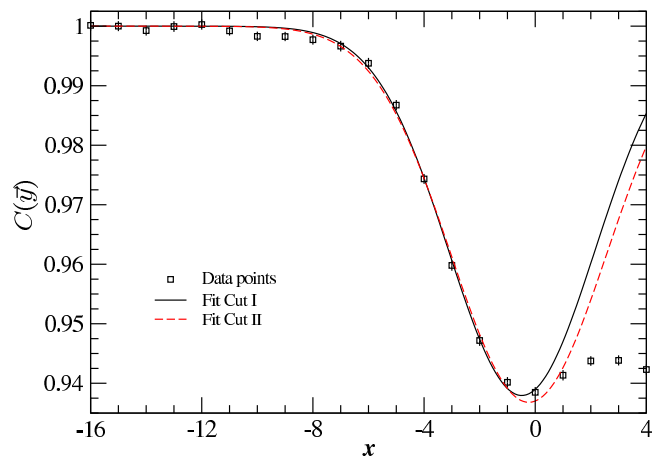}
\caption{Example fit of a Gaussian to the flux-tube node as per Table
  \ref{nodes-fit}.  The data points are for the seventh Y shape with
  30 sweeps of smearing as in the lower part of Fig.~\ref{nodes}.}
\label{fit-node-ys30}
\end{figure}

A fit using the left-hand side of the nodes with an exponential curve
of the form $1-A\, e^{-(x-x_0)^2/r^2}$ can provide a more precise
measure of the size of the node and of its actual location.  The results
have a slight dependency on the location of the cut in the data.  In
Table~\ref{nodes-fit} we show the results of the fit for the seventh
30-sweep T- and Y-shape sources for two cuts.  In cut I, only data from
$-16$ up to the minima are kept.  For cut II, data up to one lattice
unit on the right of the minima are kept. Both fits are in good agreement
with our data as is illustrated in Fig.~\ref{fit-node-ys30}.

\begin{table}
\caption{Values of the node fit parameters in the function $1-A\,
  \exp(-(x-x_0)^2/r^2)$ constrained to the left-hand side of the
  curves for the seventh 30-sweep T- and Y-shape sources in
  Fig.~\ref{nodes} for two data cuts.  Cut I considers data from
  $x=-16$ to the minima and cut II from $-16$ to one lattice unit on
  the right of the minima.  $r$ is reported in both lattice units (LU)
  and fm.}
\label{nodes-fit}
\begin{ruledtabular}
\begin{tabular}{ccccc}
Shape      & $A$        & $x_0$ (LU) & $r$ (LU)   & $r$ (fm)    \\
\noalign{\smallskip}
\hline
\noalign{\smallskip}
T (Cut I)  & $0.061(1)$ & $+1.6(1)$  & $3.7(1)$   & $0.45(1)$   \\
T (Cut II) & $0.062(1)$ & $+1.8(1)$  & $3.9(1)$   & $0.48(1)$   \\
Y (Cut I)  & $0.062(1)$ & $-0.5(1)$  & $3.7(1)$   & $0.45(1)$   \\
Y (Cut II) & $0.063(1)$ & $-0.2(1)$  & $4.0(1)$   & $0.49(1)$   \\
\end{tabular}
\end{ruledtabular}
\end{table}

Our best estimate of the ground-state node radius from the 30-sweep
Y-shape source is $r = 0.47(2)$ fm, where the uncertainty includes
both the statistical error and the systematic uncertainty associated
with the data cut.  This value is 24(9)\% larger than the flux-tube
radius of $r = 0.38(3)$ fm.  In addition, we observe a 15(3)\%
increase in the amplitude of vacuum-action expulsion, $A$, in the node
of the flux-tube relative to the expulsion observed in the flux tube
itself, for the seventh quark separation.  The amplitude of the
relative action expulsion in the node is reasonably uniform for the
various quark separations at $A = 0.062(2)$.  Here the uncertainty
reflects statistical and systematic uncertainties associated with the
data cuts and various quark separations.

In these results $r$ is linked to the radius of the node which
approaches 0.5 fm, a little larger than the flux-tube radius found
earlier for the 30-sweep smeared sources.
$x_0$ is indicative of the precise location of the node and
Table~\ref{nodes-fit} provides $x_0 = 1.7(1)$ LU for the T
shape, consistent with the results of Fig.~\ref{pot-ls-s30} and in
accord with the contour plot of Fig.~\ref{Tshape7-s30}.

For the Y shape, Table~\ref{nodes-fit} shows the node is very close to
the origin defined at the node of the source.  The movement of the
observed node position to the left is consistent with movement towards
the location of the Fermat point.  In this case the Fermat point is
located at $x = -0.54$ and $x_0$ for Cut I is in agreement with this
value.

On a final note, we explore the Euclidean time evolution of our
results for the 30-sweep smeared Y-shape source, which we have
established to access the ground state potential in a quantitative
manner.  The early plateau behavior displayed in the top plot of
Fig.~\ref{plat-s30} for the static quark potential suggests that there
should be little to no Euclidean time dependence in our observations.
Indeed this is the case for the shape of the observed vacuum action
suppression.  As emphasized in Sec.~\ref{results10}, an examination of
normalized contour plots reveals no discernible $\tau$ dependence in
the flux-tube distribution.

\begin{figure}
\centering\includegraphics[clip=true]{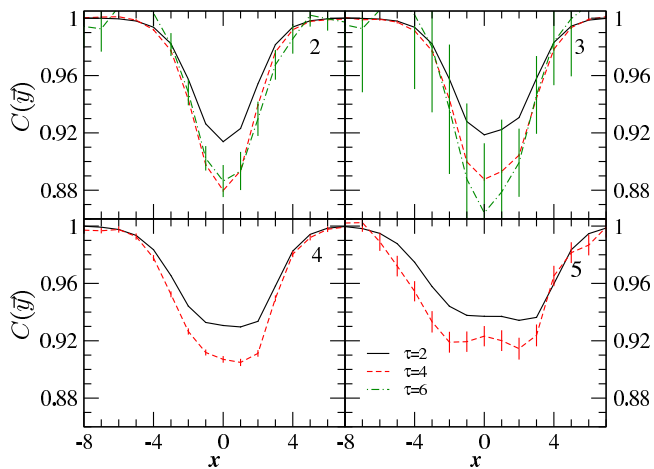}
\caption{Value of $C(x,0,0)$ for various Y-shape sources of
  Table~\ref{tab:coord} where the Euclidean time dependence of the
  vacuum-action suppression can be examined.  Here, $x$ is along the
  long direction of the quark plane ($z = 0$) and with $y=0$ the
  centre of the node is encountered.  Results are for the 30-sweep
  smeared sources.}
\label{tauDep}
\end{figure}

However, the depth of the relative vacuum-action suppression does
indeed display a $\tau$ dependence which is associated with the use of
a highly-improved lattice action density composed of 4-sweep
APE-smeared links.  Figure~\ref{tauDep} displays curves similar to
that reported in Fig.~\ref{nodes}, but this time for the second
through fifth quark separations listed in Table~\ref{tab:coord}.
Results for $\tau = 2$, 4, and 6 are illustrated for shapes 2 and 3
while the signal quality allows us to consider only $\tau = 2$ and 4
for shapes 4 and 5.  Here the Y-shape source is considered.

As the curves of Fig.~\ref{tauDep} pass directly through the centre of
the node and the flux-tube to the right of the node, the amplitude of
the relative vacuum-action suppression, $A$, may be simply read off
the plot.  While there is substantial enhancement of $A$ in going from
$\tau = 2$ to 4, the value for $A$ saturates at $\tau = 6$.

Hence, the effect of further Euclidean time evolution is to enhance the
amplitude, $A$, of the relative vacuum-action expulsion.  The
enhancement factor is uniform over the various separations at 1.3(1).
Applying this correction to our earlier results for the seventh
quark separation where the node and flux tube are well separated, we
find  $A = 0.081(7)$ and 0.072(6) for the node and flux-tube
respectively.

\section{CONCLUSIONS}
\label{conclusions}

We have studied extensively correlations of the vacuum-action density
with Wilson loops describing the three-quark static-baryon system as
defined in Eq.~(\ref{correl}).  A high-statistics approach based on
the translational and rotational symmetry of the four-dimensional
lattice volume is adopted to avoid the need for gauge-dependent
smoothing techniques.

The correlations indicate that the vacuum-action density is suppressed
in the interior and surrounding volume in accord with previous
findings in the quark-antiquark case.  ``Flux tubes'' represent the
expulsion or suppression of vacuum gluon-field fluctuations.

For small quark separations, the ground state is accessed easily
through the use of spatially-smeared APE links in the construction of
the source and sink of the three-quark system.  As illustrated in
Fig.~\ref{Ytube3}, the vacuum-action suppression is quite spherical
just outside the quark positions, and is strongest in the very centre
of the three-quark system.  As such, there is no evidence of the
formation of a $\Delta$-shape flux-tube (empty triangle) distribution.
% perhaps due to the broad nature of the flux-tube radii found in our
% study at larger quark separations.
% On the other hand, 

As the quarks become more separated the vacuum expulsion takes the
shape of a solid or filled triangle.  We conclude that at small quark
separations \textit{i.e.} for $\langle r_s \rangle \leq 0.5$ fm
Y-shape flux-tube formation does not occur.  As illustrated in
Figs.~\ref{Ttube3}, \ref{Ytube3} and \ref{Ytube4} a localized junction
of flux tubes is not observed.
% This discovery
% supports the findings of
% Refs.~\cite{Takahashi:2002bw,Alexandrou:2002sn,deForcrand:2005vv}
% based on analyses of the static baryon potential where the two-body 
% interaction 
% %and therefore the $\Delta$ law, 
% dominates.

For quark separations where the quarks are more than 0.5 fm from the
system centre, flux-tube formation is observed, in support of
Refs.~\cite{Takahashi:2002bw,Alexandrou:2002sn,%
deForcrand:2005vv,Ichie:2002mi}.  Here, the action distribution shows
some dependence on the source due to the limited Euclidean time
evolution that is possible.  Through a study of 8 T and Y shapes and
11 L shapes of varying sizes, we are able to quantitatively
demonstrate that the observed potential is a linear function of the
length of string required to connect the quark positions to the
observed position of the node connecting the flux tubes.  This is
illustrated in Figs.~\ref{pot-ls} and \ref{pot-ls-s30} for the 10- and
30-sweep sources respectively.  As such, the ground state potential at
large quark separations is the flux-tube distribution which minimizes
this length.  In this case the node will be located at the Fermat
point.

We have shown that T-shape sources have a flux-tube distribution which
relaxes towards a Y-shape distribution as the ground state of the
static quark potential is accessed.  This is illustrated in
Fig.~\ref{Tshape7-s30} with the corresponding effective potentials
presented in Fig.~\ref{ysmear-comp}.  While the formation of a Y-shape
flux-tube has been observed before (see \cite{Ichie:2002mi} for
example), to the best of our knowledge this is the first time such a
structure has been observed in a gauge invariant manner and observed to
emerge from sources that do not have a Y-shape bias.

We have successfully exposed the flux-tube distribution of the
ground-state static-quark baryon system at large separations.
Figures~\ref{Ytube7-s30} and \ref{Yshape7-s30} provide the best
illustrations of this flux-tube distribution.  Animations for a
variety of quark separations are available on the web \cite{webAnim}.

Finally the characteristic sizes of the flux-tube and node have been
quantified.  We find the ground state flux-tube radius to be 0.38(3)
fm with vacuum-field fluctuations suppressed by 7.2(6)\%.  The node
connecting the flux tubes is 24(9)\% larger at 0.47(2) fm with a
15(3)\% larger suppression of the vacuum action at 8.1(7)\%.

\bigskip

\begin{acknowledgments}
We thank Philippe de Forcrand for beneficial discussions.
This work has been done using the Helix supercomputer on the Albany
campus of Massey University and supercomputing resources from the
Australian Partnership for Advanced Computing (APAC) and the South
Australian Partnership for Advanced Computing (SAPAC).  The 3-D
realisations have been rendered using OpenDX
(http://www.opendx.org). The 2D plots and curve fitting have been done
using Grace (http://plasma-gate.weizmann.ac.il/Grace/).
\end{acknowledgments}


\begin{thebibliography}{9}

%First attempt
%\cite{Flower:1986ru}
\bibitem{Flower:1986ru}
  J.~Flower,
  %``BARYONS ON THE LATTICE. 1. ROTATIONAL SYMMETRY,''
  Nucl.\ Phys.\  B {\bf 289}, 484 (1987).
  %%CITATION = NUPHA,B289,484;%%

%Early studies of the potential

%\cite{Takahashi:2002bw}
\bibitem{Takahashi:2002bw}
  T.~T.~Takahashi, H.~Matsufuru, Y.~Nemoto and H.~Suganuma,
  Phys. Rev. Lett. \textbf{86}, 18 (2001).
  T.~T.~Takahashi, H.~Suganuma, Y.~Nemoto and H.~Matsufuru,
  %``Detailed analysis of the three quark potential in SU(3) lattice QCD,''
  Phys.\ Rev.\ D {\bf 65}, 114509 (2002)
  [arXiv:hep-lat/0204011].
  %%CITATION = HEP-LAT 0204011;%%

%\cite{Alexandrou:2001ip}
\bibitem{Alexandrou:2001ip}
  C.~Alexandrou, P.~De Forcrand and A.~Tsapalis,
  %``The static three-quark SU(3) and four-quark SU(4) potentials,''
  Phys.\ Rev.\ D {\bf 65}, 054503 (2002)
  [arXiv:hep-lat/0107006].
  %%CITATION = HEP-LAT 0107006;%%

%\cite{Alexandrou:2002sn}
\bibitem{Alexandrou:2002sn}
  C.~Alexandrou, P.~de Forcrand and O.~Jahn,
  %``The ground state of three quarks,''
  Nucl.\ Phys.\ Proc.\ Suppl.\  {\bf 119}, 667 (2003)
  [arXiv:hep-lat/0209062].
  %%CITATION = HEP-LAT 0209062;%%

%\cite{deForcrand:2005vv}
\bibitem{deForcrand:2005vv}
  Ph.~de Forcrand and O.~Jahn,
  %``The baryon static potential from lattice QCD,''
  Nucl.\ Phys.\ A {\bf 755}, 475 (2005)
  [arXiv:hep-ph/0502039].
  %%CITATION = HEP-PH 0502039;%%

%\cite{Ichie:2002mi}
\bibitem{Ichie:2002mi}
  H.~Ichie, V.~Bornyakov, T.~Streuer and G.~Schierholz,
  %``The flux distribution of the three quark system in SU(3),''
  Nucl.\ Phys.\ Proc.\ Suppl.\  {\bf 119}, 751 (2003)
  [arXiv:hep-lat/0212024].
  %%CITATION = HEP-LAT 0212024;%%
V.~G.~Bornyakov {\it et al.}  [DIK Collaboration],
  %``Baryonic flux in quenched and two-flavor dynamical QCD,''
  Phys.\ Rev.\ D {\bf 70}, 054506 (2004)
  [arXiv:hep-lat/0401026].
  %%CITATION = HEP-LAT 0401026;%%

%\cite{Cornwall:2006}
\bibitem{Cornwall:2006}
  J.~M.~Cornwall, 
  %"On the centre-vortex baryonic area law,"
  Phys.\ Rev.\ D {\bf 69}, 065013 (2004)
  [arXiv:hep-th/0305101].

%Shape dependence
%\cite{Okiharu:2003vt}
\bibitem{Okiharu:2003vt}
  F.~Okiharu and R.~M.~Woloshyn,
  %``A study of colour field distributions in the baryon,''
  Nucl.\ Phys.\ Proc.\ Suppl.\  {\bf 129}, 745 (2004)
  [arXiv:hep-lat/0310007].
  %%CITATION = HEP-LAT 0310007;%%

%\cite{Bissey:2005sk}
\bibitem{Bissey:2005sk}
  F.~Bissey {\it et al.},
  %``Gluon field distribution in baryons,''
  Nucl.\ Phys.\ Proc.\ Suppl.\  {\bf 141}, 22 (2005)
  [arXiv:hep-lat/0501004].
  %%CITATION = HEP-LAT 0501004;%%

%Mesonic methods for flux distribution
%\cite{Sommer:1987uz}
\bibitem{Sommer:1987uz}
  R.~Sommer,
  %``Chromoflux Distribution In Lattice QCD,''
  Nucl.\ Phys.\ B {\bf 291}, 673 (1987).
  %%CITATION = NUPHA,B291,673;%%

%\cite{Bali:1994de}
\bibitem{Bali:1994de}
  G.~S.~Bali, K.~Schilling and C.~Schlichter,
  %``Observing long color flux tubes in SU(2) lattice gauge theory,''
  Phys.\ Rev.\ D {\bf 51}, 5165 (1995)
  [arXiv:hep-lat/9409005].
  %%CITATION = HEP-LAT 9409005;%%

%\cite{Haymaker:1994fm}
\bibitem{Haymaker:1994fm}
  R.~W.~Haymaker, V.~Singh, Y.~C.~Peng and J.~Wosiek,
  %``Distribution of the color fields around static quarks: Flux tube
  %profiles,''
  Phys.\ Rev.\ D {\bf 53}, 389 (1996)
  [arXiv:hep-lat/9406021].
  %%CITATION = HEP-LAT 9406021;%%

\bibitem{APE87}
  M.~Falcioni, M.~L.~Paciello, G.~Parisi and B.~Taglienti,
  %``Again On SU(3) Glueball Mass,''
  Nucl.\ Phys.\ B {\bf 251}, 624 (1985).
  %%CITATION = NUPHA,B251,624;%%
  M.~Albanese {\it et al.}  [APE Collaboration],
  %``Glueball Masses And String Tension In Lattice QCD,''
  Phys.\ Lett.\ B {\bf 192}, 163 (1987).
  %%CITATION = PHLTA,B192,163;%%

%\cite{Bilson-Thompson:2002jk}
\bibitem{Bilson-Thompson:2002jk}
S.~O.~Bilson-Thompson, D.~B.~Leinweber and A.~G.~Williams,
%``Highly-improved lattice field-strength tensor,''
Annals Phys.\  {\bf 304}, 1 (2003)
[hep-lat/0203008].
%%CITATION = HEP-LAT 0203008;%%

%\cite{Luscher:1984xn}
\bibitem{Luscher:1984xn}
M.~Luscher and P.~Weisz,
%``On-Shell Improved Lattice Gauge Theories,''
Commun.\ Math.\ Phys.\  {\bf 97}, 59 (1985)
[Erratum-ibid.\  {\bf 98}, 433 (1985)].
%%CITATION = CMPHA,97,59;%%

%\cite{Montvay:1994bk}
\bibitem{Montvay:1994bk}
I. Montvay and G. M\"unster, ``Quantum Fields on a Lattice''
(Cambridge, 1994) 389.

%\cite{NRC92}
\bibitem{NRC92}
  W.~H.~Press, B.~P.~Flannery, S.~A.~Teukolsky and W.~T.~Vetterling,
  "Numerical Recipes in C"  (Cambridge University Press, 1992), 681
  and references therein. 

\bibitem{webAnim}
www.physics.adelaide.edu.au/theory/staff/leinweber/\hfill\break
VisualQCD/Nobel/ 

\end{thebibliography}
\end{document}